\documentclass[twocolumn]{arxivemptystyle}
\usepackage{bm}
\usepackage{subfig}
\usepackage{setspace}
\usepackage{amsmath,amsfonts,amssymb}
\usepackage[colorlinks=true,
            linkcolor=myred,
            urlcolor=myred,
            citecolor=myblue]{hyperref}

\newcommand{\erdosreyni}{Erd\H{o}s-Réyni}

\newcommand{\auc}{AUC}
\newcommand{\realworld}{real-world} 
\newcommand{\dtpop}{\Delta t}

\title{Detecting structural perturbations from time series with deep learning}
\newcommand{\affilications}{$^1$ D\'epartement de Physique, de G\'enie Physique, et d'Optique, Universit\'e Laval, Qu\'ebec (Qu{\'e}bec), Canada G1V 0A6\\~~Centre interdisciplinaire en modélisation mathématique de l’Université Laval, Québec, G1V 0A6, Canada\\$^*$ To whom correspondence should be addressed. E-mail: \texttt{edward.laurence.1@ulaval.ca}}
\author{Edward Laurence$^{1,*}$, Charles Murphy$^1$, Guillaume St-Onge$^1$, Xavier Roy-Pomerleau$^1$, and Vincent Thibeault$^1$}
\date{\today}

\begin{document}

\maketitle
\begin{spacing}{0.9}{\noindent \sf \bfseries \textbf{Small disturbances can trigger functional breakdowns in complex systems. A challenging task is to infer the structural cause of a disturbance in a networked system, soon enough to prevent a catastrophe. We present a graph neural network approach, borrowed from the deep learning paradigm, to infer structural perturbations from functional time series. We show our data-driven approach outperforms typical reconstruction methods while meeting the accuracy of Bayesian inference. We validate the versatility and performance of our approach with epidemic spreading, population dynamics, and neural dynamics, on various network structures: random networks, scale-free networks, 25 real food-web systems, and the C. Elegans connectome. Moreover, we report that our approach is robust to data corruption. This work uncovers a practical avenue to study the resilience of real-world complex systems.}}
\end{spacing}
{\noindent \scriptsize self-supervised learning $\vert$ graph neural networks $\vert$ complex networks $\vert$ nonlinear dynamics}\\
\vspace{0.15cm}

Complex systems can shift abruptly and irreversibly into pathological states following a disturbance. Mass extinction, stock market crash, lake eutrophication are concrete examples of such catastrophes \cite{hoegh2007coral,may2008ecology,schindler2006recent}. For systems with a clear underlying network structure, stresses may take the form of structural defects such as removing nodes or edges. These perturbations bring the system closer to a tipping-point until an obvious dynamical shift is observed \cite{scheffer2010foreseeing}. To prevent breakdowns to occur, early detection is essential but is notoriously difficult \cite{lenton2011early}. Indeed, the effect of the disturbance can be negligible and comparable to the system's noise \cite{scheffer2009early}. Even more difficult than early detection of catastrophes is the challenge of \textit{inferring} the removed edges/nodes.

Given observations of a network dynamics, the issue of identifying structural defects has remained largely unexplored. Most closely related are numerous studies that tackle the problem of detecting early-warning signals of functional transitions \cite{dakos2015resilience,scheffer2009early,clements2019early,wissel1984universal}. Mostly grouped under the critical slowing down paradigm, these methods are far from universal as many have reported erroneous detections or failed to forecast an eventual catastrophe \cite{boerlijst2013catastrophic,wilkat2019no,jager2019systematically,kefi2013early}. Moreover, few early-warning signals take into account the structure of interactions, even though data may be available, so the structural cause of the disturbance cannot be inferred. Other approaches coming from the study of dynamical complex networks investigate the functional effect of removing edges \cite{gao2016universal,jiang2018predicting,laurence2019spectral}. However, they are restricted to specific dynamical models which are generally inadequate to describe the rich behavior of real complex systems.

Reconstruction methods have been frequently used, especially in neuroscience, to infer a functional structure from complex time series \cite{seth2015granger,barzel2013network,sheikhattar2018extracting,roebroeck2005mapping,chen2006frequency,seth2015granger,bressler2011wiener}. Most of these methods shine by their simplicity as they usually do not require any parameter fitting and can readily be applied to almost any type of dynamics. While being originally designed to reconstruct whole networks, they nevertheless seem a natural fit to identify structural defects. Yet, the precision of the reconstructed structure is often limited when the methods are applied on \realworld{} datasets. They tend to detect a large variety of functional (dense) relationships between the nodes, which are only indirectly related to the (sparse) structure \cite{feizi2013network,park2013structural}. Other reconstruction approaches rely on assumptions of dynamical \cite{peixoto2019network,shandilya2011inferring} or structural models \cite{young2019reconstruction,newman2018estimating,pan2016predicting}. Their main assets are their high accuracy on toy models with ground-truth data and their quantification of the confidence interval of the reconstructed structure. However, these approaches are limited to specific instances where the model assumptions are reasonable \cite{brugere2018network}, which limit their scope for \realworld{} applications.

In this paper, we directly tackle the issue of inferring structural perturbations by introducing a method based on graph neural networks oriented toward \realworld{} applications. We achieve this by training a deep learning model to forecast complex dynamics after being trained on time series of activity with a given original network. Our approach is self-supervised and predicts structural defects without prior knowledge of the nature of the perturbations, nor of the dynamical process that generated the time series. It can thus be applied to various dynamics and networks with minimal modifications. We benchmark its performance on epidemiological, population, and neural dynamics over various synthetic and real networks. We show that it outperforms standard reconstruction methods while achieving a high level of precision, and proves to be greatly robust to noisy datasets.

Our work provides a novel method to study and monitor stressed complex networks. Beyond paving the way of bringing deep learning frameworks into the study of dynamical networks, our versatile approach can be used to address concrete problems of \realworld{} systems. To name a few, it includes monitoring ecological systems \cite{pires2017rewilding}, documenting temporal evolution of gene regulatory networks \cite{hecker2009gene}, detecting leaks in water flow networks \cite{perez2009pressure}, and evaluating pathological structures of brains disorders \cite{lynall2010functional}.

\begin{figure*}	  
	\centering
	\centering\includegraphics[width=0.31\linewidth]{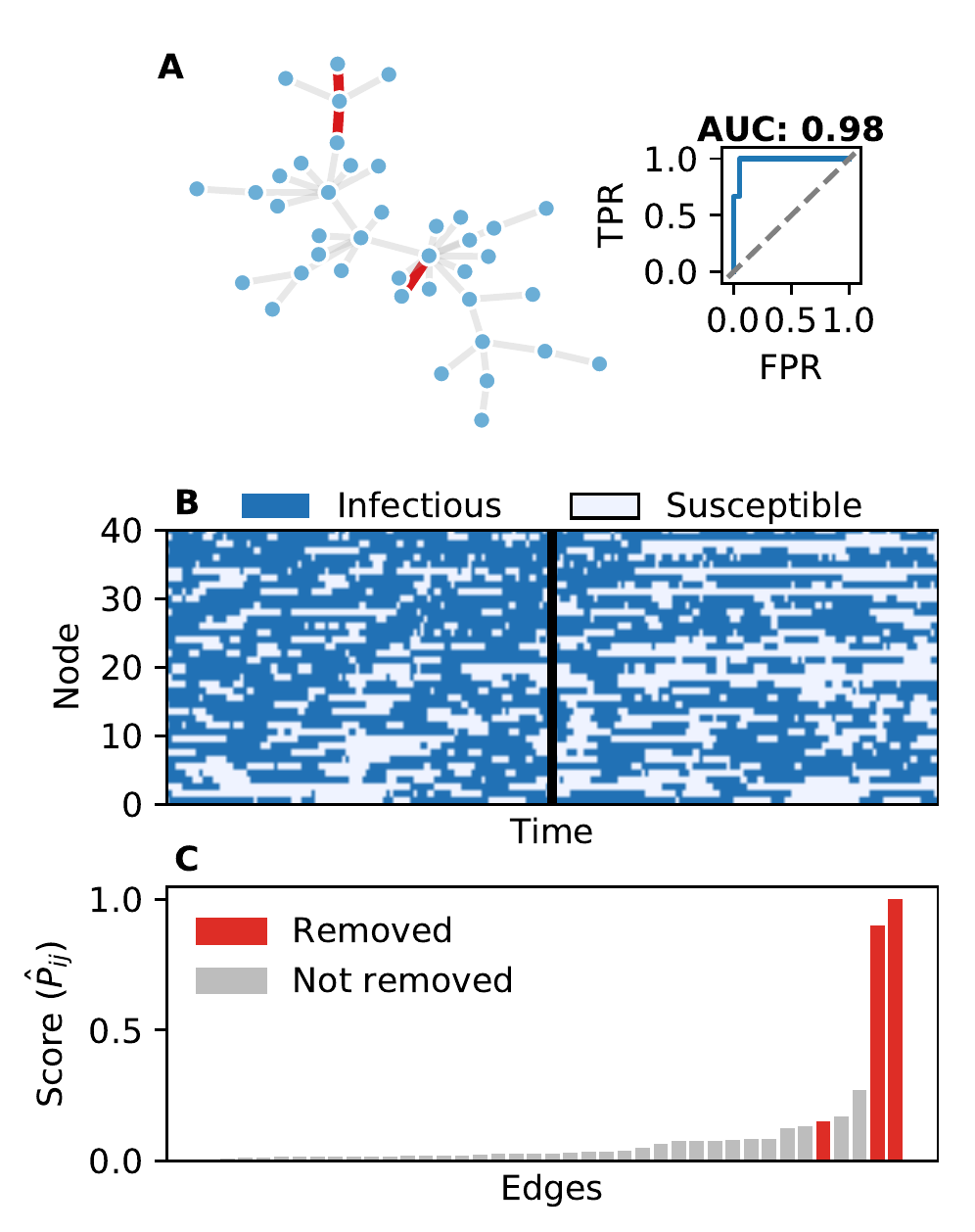}\quad
    \centering\includegraphics[width=0.32\linewidth]{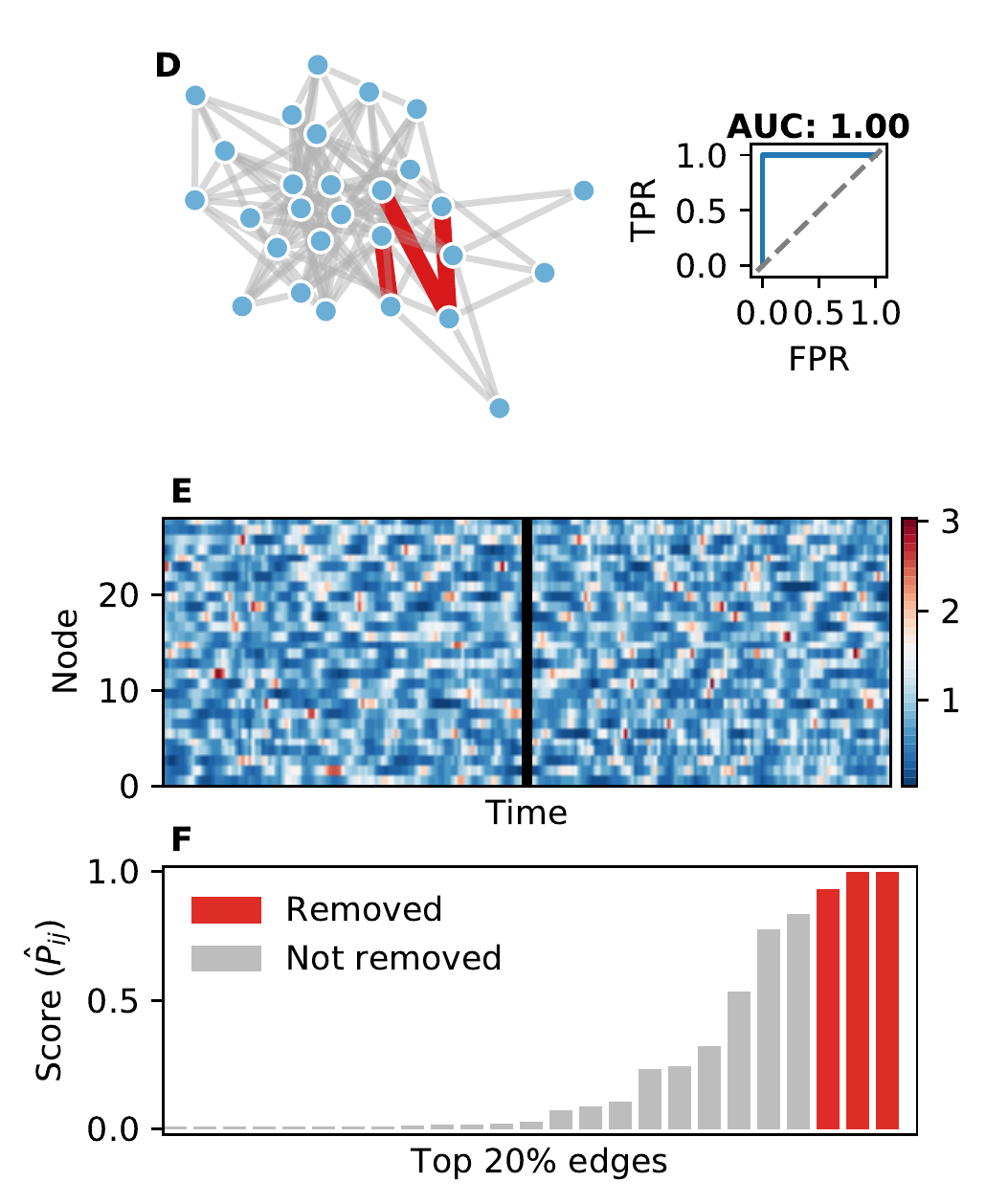}\quad
    \centering\includegraphics[width=0.31\linewidth]{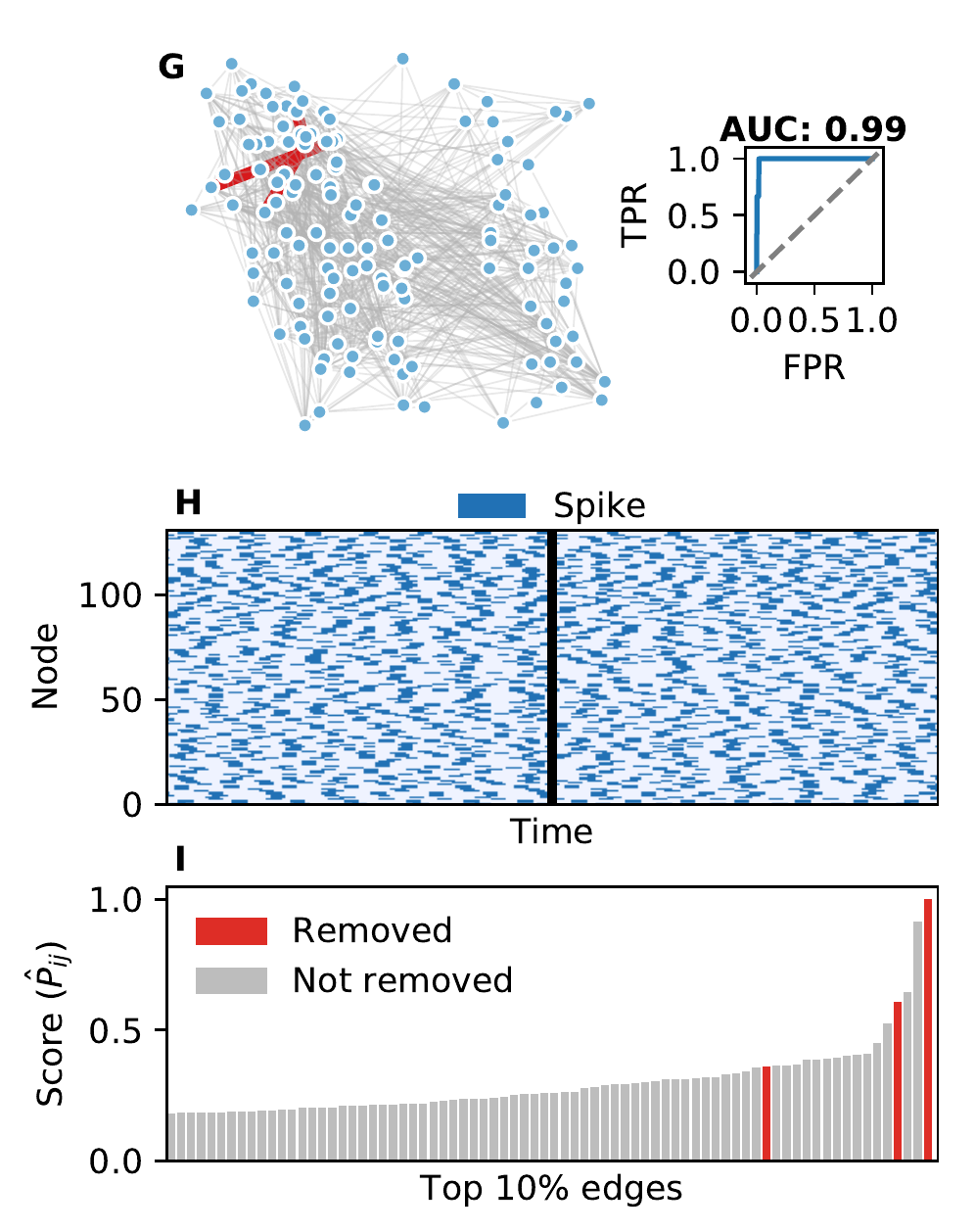}%
	\caption{Examples of GNN predictions and time series for epidemic spreading on a scale-free network (A-B-C), population dynamics on a real ecological network \cite{angelini2011ecosystem} (D-E-F), and neural dynamics on a C.Elegans connectome (G-H-I). Halfway through the time series (black lines in B-E-H), three random edges, shown by the thick red edges in (A-D-G), are removed from the original networks displayed in (A-D-G). The bars of (C-F-I) show the predicted scores by the GNN model on individual edges with true removed edges colored in red. The ROC curves of the GNN predictions are shown with the corresponding \auc{}. }
	\label{Fig:example}
\end{figure*}
 
\section{Context}
Consider a graph $G$ composed of $N$ nodes and a set $\mathcal{E}$ of $M=|\mathcal{E}|$ edges. We denote the adjacency matrix by $\bm{A}$ whose element $a_{ij}=1$ if there exists an interaction from node $j$ to node $i$ and $a_{ij} = 0$ otherwise. We consider $T$ discrete observations of the nodes activity represented by a time series $\bm{X}=\lbrace \bm{x}(t)\in \mathbb{R}^{N}\rbrace_{t=1...T}$, where the element $x_i(t)=[\bm{x}(t)]_i$ is the activity of node $i$ at time $t$. The time series is generated from a hidden and potentionally stochastic dynamical mechanism,
\begin{equation}
	\bm{X} = \mathcal{M}[\bm{A}, \bm{X}_0, \bm{\theta}],\label{C4:Eq:dynamecanism}
\end{equation}
where $\bm{X}_0$ is the initial state of the system and $\bm{\theta}$ are unknown parameters. The dynamics can take any form respecting the condition that the time evolution of a node activity depends on itself and the activity of its neighbors only. It ensures that the adjacency matrix truly governs the interactions.

We consider the following scenario: For $t<\tau$, the dynamics governed by \eqref{C4:Eq:dynamecanism} is taking place on the original graph whose edges are $\mathcal{E}$. At $t=\tau>0$, the graph is perturbed by removing a set $d\mathcal{E}\subset\mathcal{E}$ of edges. The effect of the perturbation is a shift in the adjacency matrix $\bm{A}'=\bm{A}-\bm{P}$ where $\bm{P}$ is called \textit{the perturbation matrix}. For $t\geq\tau$, the dynamics takes place on the perturbed graph of adjacency matrix $\bm{A}'$. Note that the change point $\tau$ is only introduced to simplify the notation and is assumed to be known. Our analysis could have been done equivalently with the more general scheme of having two distinct and disconnected time series, the dynamics occurring on the original graph and the dynamics on the perturbed graph. 

We now address the task of inferring the perturbation matrix $\bm{P}$ relying on the initial adjacency matrix $\bm{A}$, time series $\bm{X}$, and the moment of perturbation $\tau$. Three examples illustrating the task are displayed in Fig.~\ref{Fig:example}. Reasonably, this problem can be solved by inferring the perturbed structure, $\bm{A}'\approx \hat{\bm{A}}'=f(\bm{X})$ from the observation of the nodes activity. The estimated perturbation matrix $\hat{\bm{P}}$ is composed of the original edges missing from the inferred structure, i.e., $\hat{\bm{P}}=\bm{A}-\hat{\bm{A}}'$. However, reconstructing networks from time series is a notoriously hard problem and, yet, no universal method exists \cite{marbach2012wisdom}. Assuming that the structure $\bm{A}$ and nodes' activity before the perturbation $t<\tau$ are known, properly incorporating these features into the reconstruction process becomes a critical facet of the inference methods, i.e., $ \hat{\bm{A}}'=f(\bm{A}, \bm{X})$ 
 
\subsection{Graph Neural Networks}
\newcommand{\pgnn}{\hat{\bm{h}}_{{\text{GNN}}}}
We introduce a Graph Neural Networks (GNN) to leverage the given structural information. In recent years, GNN have been developed to be used on structural datasets \cite{xu2018powerful} with graph-based tasks \cite{kipf2016semi,hamilton2017inductive}. The developed model is trained on the self-supervised task of forecasting the nodes activity while optimizing on numerous internal parameters. 
More precisely, our GNN model forecasts a node activity from previous dynamical states of the neighbors,
\begin{equation} 
	\hat{\bm{x}}(t+1) = \begin{cases}
   \text{GNN}(\bm{A},\,\bm{X}^{(a)}(t),\,\bm{\Lambda}),& \text{if } t< \tau,\\
     \text{GNN}(\bm{A}-\sigma(\pgnn),\,\bm{X}^{(a)}(t),\,\bm{\Lambda}),              & \text{if }t\geq \tau
\end{cases}\label{C4:Eq:GNN}
\end{equation}
where $\bm{X}^{(a)}(t)=\lbrace \bm{x}(t')\rbrace_{t'=t-a, ..., t}$ is the $a$ past steps of activity, $\hat{\bm{x}}(t+1)$ is the forecasted activities at time $t+1$, $\bm{\Lambda}$ and $\pgnn\in\mathbb{R}^{N\times N}$ are trainable parameters, and $\sigma(\cdot)$ is the sigmoid function. In \eqref{C4:Eq:GNN}, the graph of interactions is given by the structure $\bm{A}-\sigma(\pgnn)$ if the forecast is next to the perturbation $t>\tau$, and simply $\bm{A}$ otherwise. Hence, the perturbation and the dynamical mechanisms are parametrized separately using $\pgnn$ and $\bm{\Lambda}$ respectively. The inferred perturbation matrix is given by $\hat{\bm{P}}=\sigma(\pgnn)$. Note that the sigmoid function $\sigma(\cdot)$ is used to limit the perturbation amplitude between 0 and 1.

During training, examples of previous and future activities are presented to the model and the error on a loss function $L[\hat{\bm{x}}, \bm{x}]$ between the forecasted activity $\hat{\bm{x}}$ and the observed activity $\bm{x}$ is backpropagated through the model for parameters optimization [See the Materials and Methods for details]. Eventually, the model reaches a stable minimum of the loss function and the perturbation matrix can be estimated as $\hat{\bm{P}}=\sigma(\pgnn)$. Model and training details are provided in the Materials and Methods section.

By design, the GNN is inductive, which means that its predictions are independent of the target node identity and only depend on the states of the target node and its neighbors. The learned dynamical mechanism is shared among all nodes, although the neighborhood activity may differ. It also means that forecasting all $N$ node states  is equivalent to operate $N$ independent forecasts, one for each node of the graph. Hence, the number of parameters of $\bm{\Lambda}$ does not have to scale up with the number of nodes in the graph and the GNN model can be lightweight memory-wise and computationally efficient.

\section{Results}
We introduce two types of inference methods to compare with the GNN model. The first are functional reconstruction algorithms. The adjacency matrix after the perturbation $\bm{A}'$ is approximated by ad-hoc metrics such as the correlation matrix \cite{barzel2013network} or the Granger Causality \cite{granger1969investigating,bressler2011wiener,chen2006frequency,roebroeck2005mapping,seth2015granger}, and the perturbation matrix is estimated by finding the original edges missing from the inferred structure $\hat{\bm{P}}=\bm{A}-\hat{\bm{A}}'$. 

For the second type, we assume that the dynamical mechanisms are perfectly known. It cuts out the challenge of learning the dynamics to focus on detecting the perturbation. In the context of stochastic dynamics, we develop a Bayesian inference method, inspired by Ref.~\cite{peixoto2019network}, that estimates the perturbation by sampling the posterior distribution $\Pr(\bm{P}|\bm{A}, \bm{X}, \bm{\theta})$. This method is deeply advantaged compared to the GNN model as the dynamical mechanism $\mathcal{M}[\cdot, \cdot, \bm{\theta}]$ is assumed to be known and is used explicitly to compute the posterior distribution. Therefore, it serves as an idealized reference point to compare with our GNN model.

We measure the performance of the algorithms using the area under the curve (\auc{}) of the ROC curve. The \auc{} can be interpreted as the probability that the model distinguishes whether an edge is present or absent from the perturbation set $d\mathcal{E}$, independently of the threshold applied on the estimated perturbation matrix $\hat{\bm{P}}$. Note that an \auc{} equals to 0.5 is achieved with a uniform and non-informative baseline, whereas an \auc{} equals to 1 indicates that all the edges from the ground-truth perturbation have the top scores on the inferred perturbation matrix $\hat{\bm{P}}$.

\subsection{Spreading dynamics}
We evaluate the GNN model over an epidemic spreading dynamics called the susceptible-infectious-susceptible model (SIS) \cite{pastor2015epidemic}. Nodes can be either infected ($x_i=1$) or susceptible ($x_i=0$). At each time step, infected nodes can infect their susceptible neighbors with a probability $\alpha$. Infected individuals, on the other hand, become susceptible again with probability $\beta$. An example of a time series and prediction of the perturbations is given in Fig.~\ref{Fig:example}(A,B,C).

\begin{figure}
	\centering
	\includegraphics[width=\linewidth]{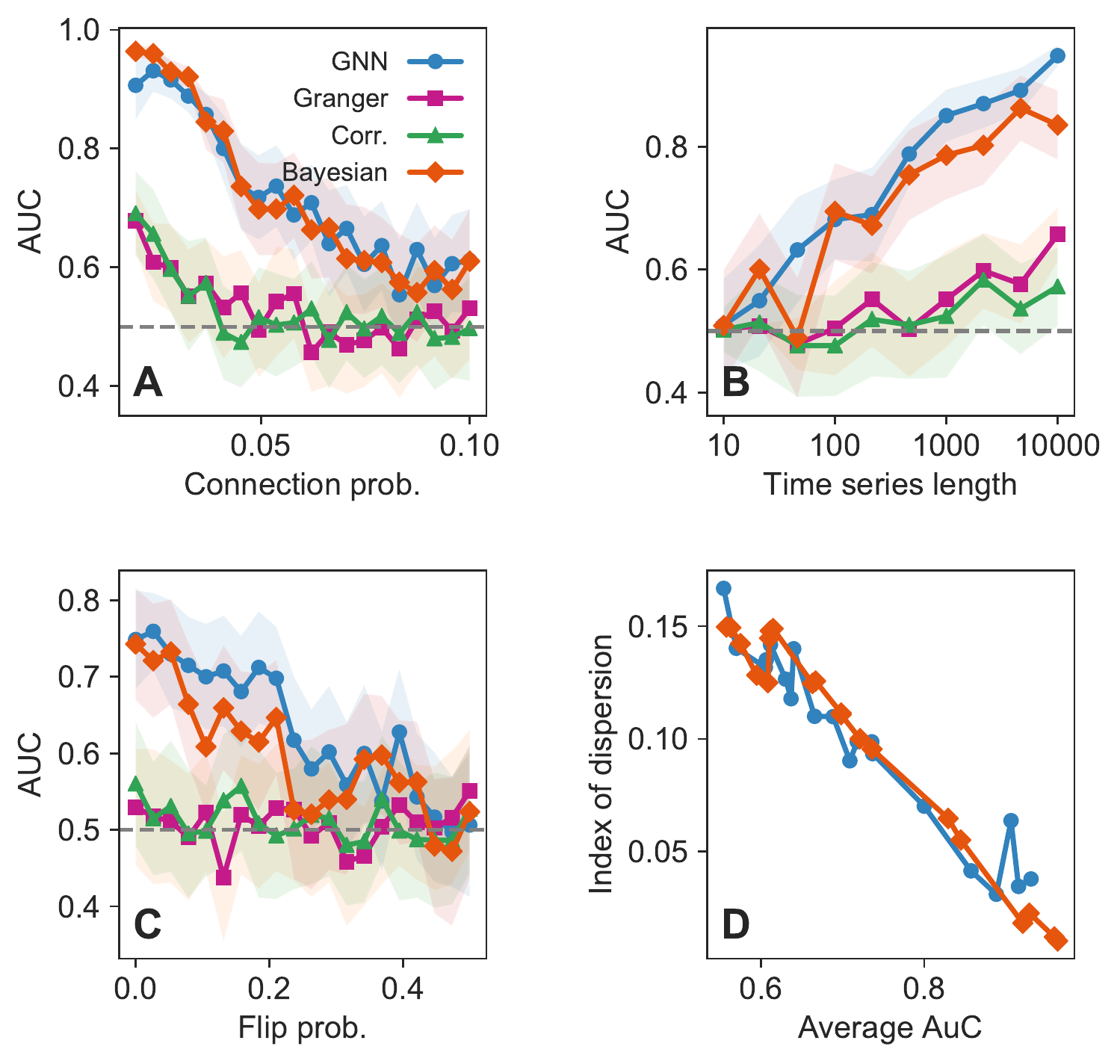}
	\caption{Area under the curve for SIS dynamics on \erdosreyni~networks. (A) With different network connection probability $p$ and $N=100,\,T=300,\,\alpha=\beta=0.1$. (B) With different time series length and $N=100,p=0.05$, and transition probabilities $\alpha=0.2,\,\beta=0.1$. (C) \auc{} under various flip probabilities with $N=100,\,p=0.05,\,T=300,\,\alpha=\beta=0.1$. Each symbol is the average over 100 independent simulations with similar configurations and standard deviation is displayed by a shaded region. The gray dashed lines are the non-informative baseline \auc{}$=0.5$. (D) Index of dispersion $D=\sigma^2/\mu$ of \auc{} for data from (A).}
	\label{Fig:SIS:Erdos}
\end{figure}

\subsubsection{Random networks}
In Fig.~\ref{Fig:SIS:Erdos}, we report the \auc{} for spreading dynamics over the \erdosreyni~random networks \cite{newman2018networks} for various parameters. It is striking that the GNN performs as well as the Bayesian model, a result that is consistent over all experiments. This may come as a surprise as the GNN has been given a more difficult task: Learning both the dynamics and the perturbation. Thus, it suggests that the GNN has simultaneously achieved the task of learning the dynamical mechanism and struggles as much as the Bayesian method on detecting the perturbation.

Another important result is that the GNN model outperforms functional reconstruction methods over all experiments. It demonstrates the importance of explicitly considering the dynamical mechanism and the prior network structure, instead of using a purely functional-based definition of the connectivity.

All methods perform best on low connectivity networks [Fig.~\ref{Fig:SIS:Erdos}(A)]. Perturbations on denser networks are harder to infer, independently of the model used. At least two phenomena justify this. First, the number of edges involved in the perturbation is fixed while the total number of edges grows with the average connectivity of the network. Hence, finding the right removed edges in an increasing number of candidates naturally lead to a more difficult task, irrespectively of the method employed. Second, the average number of infected nodes monotonically increases with the connectivity of the network [See Supplementary Information]. Therefore, the probability of being infected remains almost the same after the removal of a single edge in dense networks. Consequently, perturbations are barely detectable.

Figure~\ref{Fig:SIS:Erdos}(B) shows that increasing the length of the time series leads to better performance, especially for the GNN and Bayesian approaches. Larger time series are equivalent to having a larger dataset, hence a more robust inference. This is beneficial for the GNN in two ways since both the inference of the dynamics and of the perturbations are enhanced. Building on results of Fig.~\ref{Fig:SIS:Erdos}(A), we conclude that perturbations on dense random networks can be inferred with high confidence (\auc{}>0.9) if using sufficiently large time series, e.g., $T\sim 10^4$ entries.

The consistency of the inference can be expressed using the index of dispersion $D=\sigma^2/\mu$ of the \auc{} over experiments sharing similar settings. We observe that the dispersion decreases with the average \auc{} [Fig.~\ref{Fig:SIS:Erdos}(D)]. It means that the GNN model becomes even more consistent on similar experiments with high \auc{} outcomes. Moreover, because the dispersion coefficient is highly similar between the GNN model and the Bayesian model, it suggests that the error source is similar. We hypothesize that large \auc{} variance is produced by inherent fluctuations in the generated dataset rather than the sensitivity of the algorithms.

\subsubsection{Scale-free networks}
Networks featuring hubs that dominate the connectivity are common in nature \cite{barabasi2009scale}. We benchmark the models on tree-like scale-free networks generated from the Barabási–Albert model \cite{barabasi1999emergence}(Fig.~\ref{Fig:SIS:Bara}). In the Supplementary Information, we provide a validation of the GNN model with non-tree scale-free networks as well.

In general, the GNN model performance is much higher on scale-free networks than on random networks. This may be due to the tree-like structure of the generated networks. Perturbations tend to break up networks into disconnected components. Smaller and peripheral components tend to quickly deactivate. The removed edges may then be easily identified as the bridges between the dynamically distinct components. This could further be supported by their notable contributions in the dynamical likelihood [See Supplementary Information for details].

Fig.~\ref{Fig:SIS:Bara}(A) shows a decrease in \auc{} for the functional reconstruction methods as the infection probability increases. Based on the previous discussion, increasing the infection probability raises the contrast between the inactive and active components, which would normally help infer the perturbation. To explain this behavior, note that the correlation \eqref{C4:Eq:Correlation} between an active node $\bm{x}_i\approx \bm{1}_T$ and an inactive node $\bm{x}_j\approx \bm{0}_T$ is roughly equal to zero, but so is the correlation for pairs of active nodes. Similar arguments can be made for the Granger causality. Therefore, for high infection probability, structural information becomes hidden from functional reconstruction methods, explaining their poor performance. They perform best when the infection is sparse. It highlights the importance of using techniques that take the nature of the dynamics into consideration.

In Fig.~\ref{Fig:SIS:Bara}(B), the models are tested against increasing network size. We observe that the GNN model performance is roughly constant up to $10^3$ nodes. This may come as a surprise when recalling that the perturbation is composed of a single edge and that the model has only $300$ time steps to learn both the dynamics and the perturbation. Note that we observe a slight decrease in the performance of the Bayesian model over large networks. However, this trend is due to an insufficient number of sampling steps compared to the network size.

\begin{figure}
	\centering
	\includegraphics[width=\linewidth]{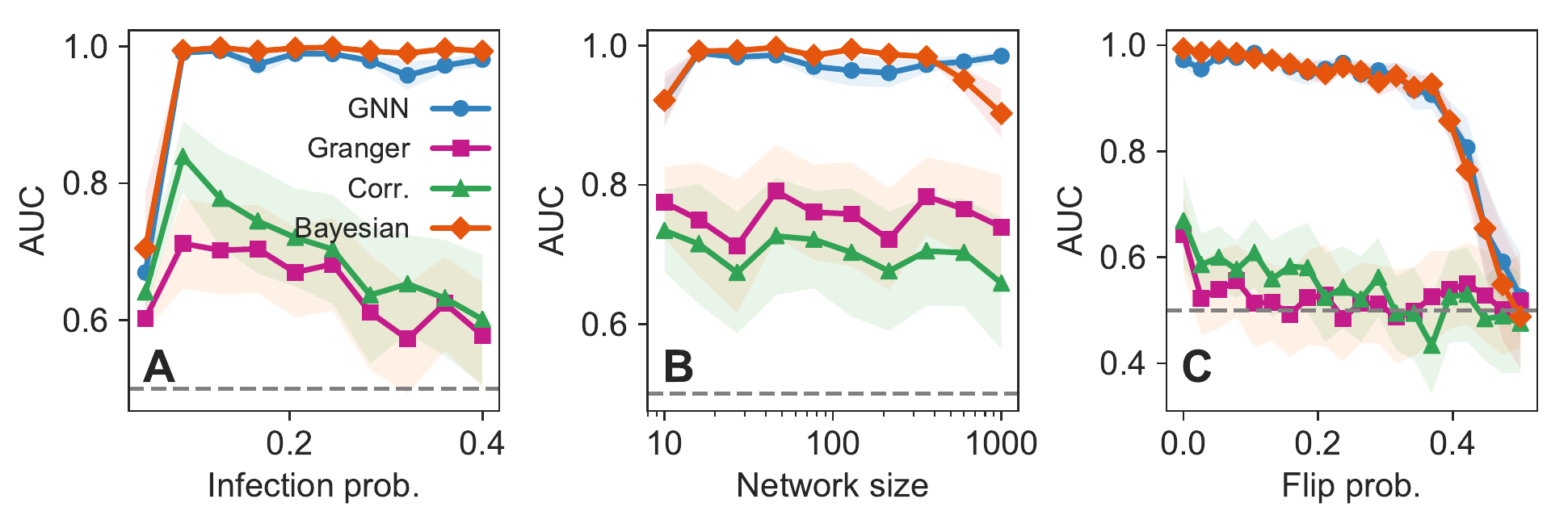}
	\caption{Area under the curve for SIS dynamics on Barabási–Albert scale-free networks. (A) Prediction over various infection probability $\alpha$ on networks of $N=100$ nodes. (B) \auc{} over increasing network size with $\alpha=0.2$. (C) \auc{} over increasing flip probability with $\alpha=0.2$. The likelihood of the Bayesian model of (C) is corrected to integrate the flip mechanism. Dots are averages over 100 simulations with similar configurations, and standard deviations are indicated by shaded regions. Epidemics spreading are simulated over $T=300$ time steps and $\beta=0.1$. Perturbations occur at $\tau=150$ and consists in removing a single random edge.}
	\label{Fig:SIS:Bara}
\end{figure}

\subsubsection{Noisy time series}
\newcommand{\pflip}{p_{\text{flip}}}
Typical real datasets are noisy \cite{mann1996robust}. Time series come as a mixture of some hidden dynamical mechanism masked by a layer of noise. We investigate the robustness to noise of the GNN model. To simulate noise, entries of the time series are flipped with probability $\pflip$, i.e., 0 goes to 1 and 1 to 0. For $\pflip=0.5$, we obtain uncorrelated time series (independent Bernoulli processes).

For the next experiments, we use a corrected Bayesian model with a likelihood function that explicitly considers the flip mechanism, including the ground-truth flip probability [See Materials and Methods]. This modification to the Bayesian model is essential, as new transitions originally forbidden become possible with the introduction of noise. Precisely, in the original Bayesian version, null values of likelihood appear and sampling of the posterior distribution becomes impossible. While the Bayesian model must be revisited, our GNN model remains unchanged, incorporating automatically the effect of noise in its predictions during training.

We compare the performance over various flip probabilities on scale-free networks [Fig.~\ref{Fig:SIS:Bara}(C)] and random networks [Fig.~\ref{Fig:SIS:Erdos}(C)]. We report that the GNN model is as precise as the corrected Bayesian model. On scale-free networks, it maintains high \auc{} over large flip probability. This performance is quite remarkable as the GNN model has to learn the dynamical mechanism from a noisy time series while inferring the missing edge. For random networks, the results show high \auc{} variances. It indicates that the noise process leads to larger fluctuations in precision, which is shared among the GNN and the Bayesian models. We again report that functional reconstruction methods are outperformed by the GNN approach. 

\begin{figure}
	\centering
	\includegraphics[width=\linewidth]{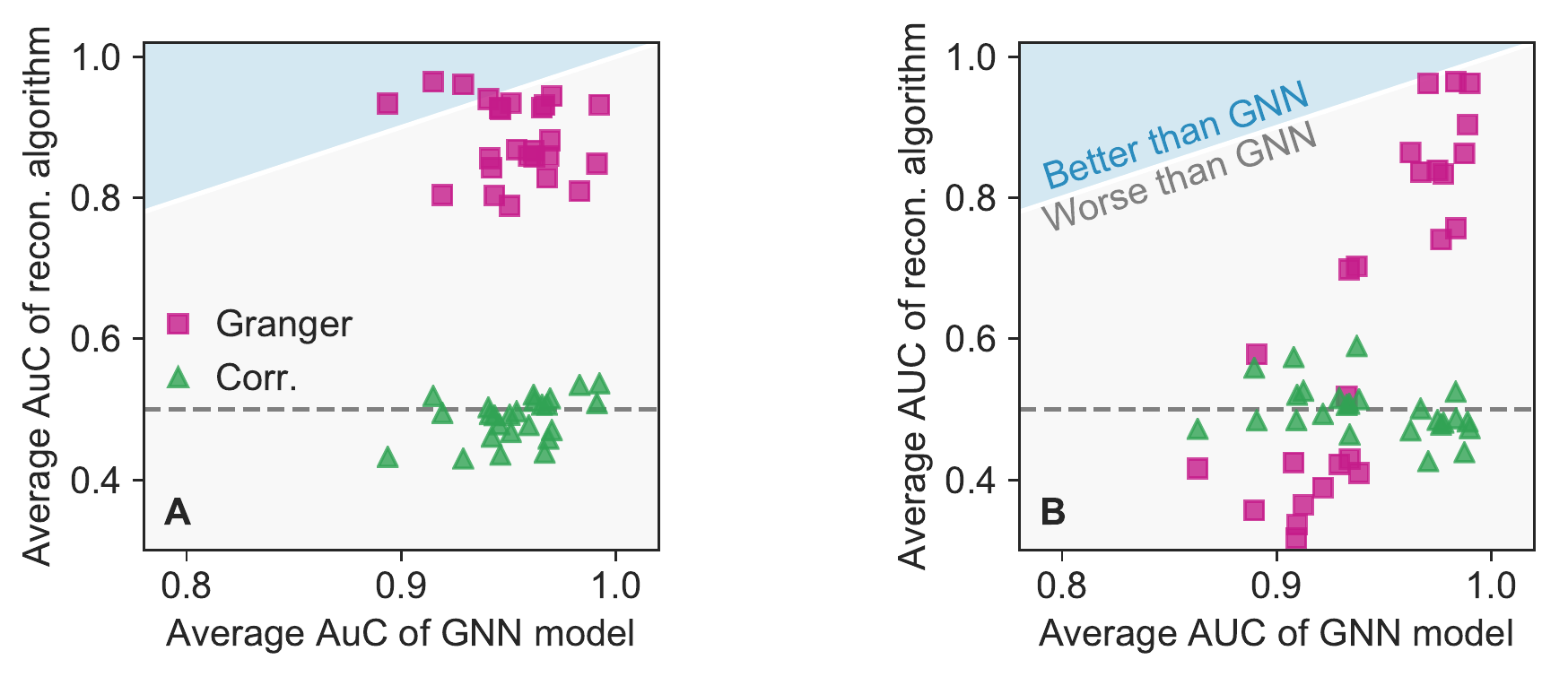}
	\caption[Average area under the curve of predictions on 25 real directed ecological networks with predator-prey dynamics.]{Average area under the curve of predictions on 25 real ecological networks with predator-prey dynamics. Each marker is associated to a specific real network. The dynamics are simulated on $T=10^3$ steps with (A) $\dtpop=1/5$ and (B) $\dtpop=1$. The functional reconstruction algorithms perform better when the dots are in the upper blue region of the line $x=y$. Random baselines are indicated with dashed gray lines. Averages are computed over 100 runs on each network. The perturbation is the removal of a single random edge at $\tau=500$.}
	\label{Fig:Lotka}
\end{figure}

\subsection{Population dynamics}
\newcommand{\tmax}{t_{\text{max}}}
Population dynamics are valuable tools to study ecological systems. Species population varies according to their predator-prey relationships. 

In these systems, each node represents a different species and its population $x_i(t)$ varies in time according to a predator-prey relationship. We benchmark our models on a simple population dynamics given by
\begin{equation}
	\dot{x}_i = c_ix_i + x_i\sum_{j=1}^N (a_{ij}-a_{ji})x_j, \label{C4:Eq:LotkaVolterra}
\end{equation}
where $c_i=c\sum_{j}(a_{ji}-a_{ij})\in \mathbb{R}$ is the fraction of the number of interactions that a species has [See Materials and Methods]. Note that introducing a species-dependent parameter $c_i$ raises the difficulty for dynamical learning as the dynamical parameters are now specific to each node. Now that the dynamics is deterministic, the Bayesian approach previously developed no longer applies. Hence, the GNN model is compared with the correlation and the Granger causality approaches on 26 real directed ecological networks [Example in Fig.~\ref{Fig:example}(D,E,F)]. We also introduce the sampling parameter $\dtpop$ that indicates the time interval between two steps in the time series. Large $\dtpop$ implies a coarser sampling of time series.

Our results show that the GNN (Granger) approach reaches an average \auc{} of 0.95 (0.88) over all simulations for $\dtpop=1/5$ and 0.94 (0.64) for $\dtpop=1$ [Fig.~\ref{Fig:Lotka}]. The GNN performance suggests that it is a valuable candidate for deterministic context, and even when the dynamical parameters of each node are dependent of their local structure.

Interestingly, the GNN model maintains a high \auc{} for all 26 real networks and large sampling steps $\dtpop$. The GNN performance is mostly unaffected by the characteristic of the network structure. 

\subsection{Neural dynamics}
We challenge the versatility of our GNN model on a neural dynamics simulated on the C.Elegans connectome. 
First, we simulate the continuous noisy theta model of \eqref{SI:Eq:thetamodel} on the directed network structure \cite{ermentrout1986parabolic}. Then, we threshold the time series to only record the spikes of activity. Doing so cuts off valuable information for learning the dynamics which greatly increase the difficulty of the inference task [Example in Fig.~\ref{Fig:example}(G,H,I)]. Also, the simulated theta model includes a random noise injection during integration.

\begin{figure}
	\centering
	\includegraphics[width=0.8\linewidth]{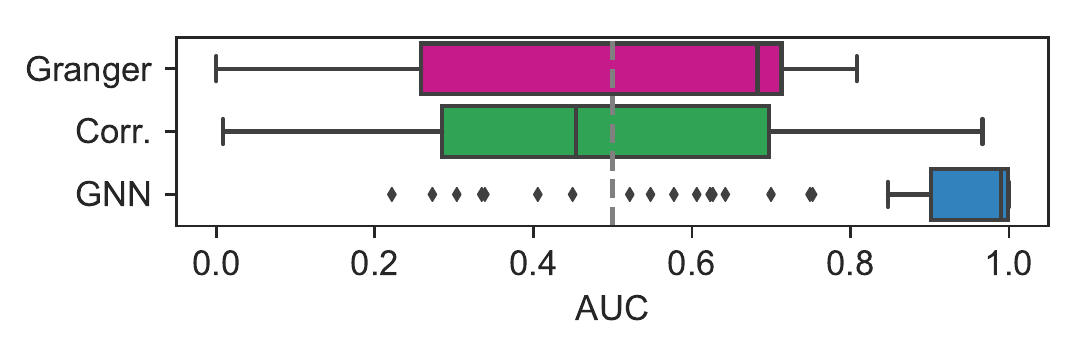}
	\caption{Area under the curve of predictions over neural dynamics on the C. Elegans connectome.  The perturbation is composed of three random edges and dynamics spans over $T=500$ time steps with $\tau=250$. Boxplots are computed over 100 simulations. Outliers (outside of 75th percentile plus 1.5 interquartile range) are indicated with markers. }
	\label{Fig:Celegans}
\end{figure}

Results indicate that the GNN model is highly capable of handling this spiking neural dynamics [Fig.~\ref{Fig:Celegans}]. Whille 17\% of the runs show \auc{} below 0.75, most of them are above the expected values from functional reconstruction algorithms. On average, the \auc{} is $0.91\pm 0.19$ with a median value at $0.99$. It demonstrates the GNN's high accuracy in the deterministic context, even in the presence of partial information. 

\section{Discussion}
Detecting perturbations of complex systems is a challenge of paramount importance, especially in modern times with the impacts of climate changes on ecosystems and our more than ever entangled societies. Yet, the lack of reliable predictive tools to infer the presence and the structural cause of disturbances is of serious concern. Without taking full advantage of the graph structure, functional effect of perturbations can be so subtle that it could remain undetected at the local and global levels.

In this paper, we have introduced a method relying on recent deep learning advances to infer structural perturbations from time series of activity. The core and original idea is the self-supervised training scheme that neither requires prior information on the perturbation strategy nor on the dynamical mechanism. Minimal modifications of the model are required to apply the method on datasets of different natures, such as continuous, noisy, or discrete time series. 

We have tested the method in three contrasting contexts: spreading, ecological, and neural dynamics. We show that our approach outperforms standard functional reconstruction methods while being comparable to Bayesian models with a priori knowledge of the dynamical mechanisms. Our results suggest that GNN are promising candidates to predict structural disturbances in practical applications.

Apart from its effectiveness, there are multiple advantages of using the GNN model. First, the optimization procedure scales, both in speed and accuracy, to large networks, and is fairly robust to the hyperparameters choice. Second, GNN models are part of an active field of research so that future improvements are expected, e.g., enhanced neural network architecture and improved gradient descent optimizer \cite{liu2019variance}. Third, GNN are versatile to different dynamics while being able to cope with large levels of noises. Finally, they are well adapted to support other types of information collected on the graph that could help the predictions.

Our study provides a new avenue for monitoring complex systems. It can be used for detecting disturbances in various classes of complex networks, ranging from ecocology to neuroscience, and paves the way for the development of targeted intervention strategies for sustainable management. 

\section{Materials and Methods}
\begin{spacing}{1}
\footnotesize
\subsubsection*{Empirical Networks} 
The ecological networks are food webs obtained from the Web of Life database (www.web-of-life.es). Supplementary information gives an overview of their structural properties. The C.Elegans connectome is obtained from Ref.~\cite{kaiser2006nonoptimal}.

\subsubsection*{Generated networks} Random networks are generated from the \erdosreyni~model: $N$ nodes are randomly connected with probability $p$ \cite{newman2018networks}. The scale-free networks are generated from the Barabási–Albert model \cite{barabasi1999emergence}. We first start with a connected pair of nodes. Then, at each step, we add a new node to the network and connect it to $m=1$ existing node chosen with probability proportional to their degrees.

\subsubsection*{Simulated dynamics}
For all experiments, we first initialize the nodes activity and run the dynamics for 100 burning steps which are then discarded. Then, we simulate the dynamics for $\tau$ steps. At $t=\tau$, we remove random edges from the original network. Then, the dynamics is continued until $t=T$. The time series $\bm{X}=\lbrace \bm{x}(t)\in \mathbb{R}^{N}\rbrace_{t=1...T}$ is then constructed from the vectors of nodes activity at each time step.

\subsubsection*{Susceptible-Infected-Susceptible dynamics (SIS)} Nodes activity is binary $\bm{x}(t) \in \lbrace 0,1\rbrace^{N}$ where $x_i(t)=1$ indicates that node $i$ is infected at time step $t$ and $x_i(t)=0$ if susceptible. At each time step, a susceptible node $i$ is infected with probability $1-(1-\alpha)^{l_i}$, where $l_i=\sum_i a_{ij}x_j$ is its number of infected neighbors and $\alpha$ is the transmission probability. Infected nodes recover with constant probability $\beta$. The nodes activity is initialized with equiprobable binary values.

\subsubsection*{Lotka-Volterra dynamics}
We have simulated mixed ecosystems from a generalized Lotka-Volterra dynamics \eqref{C4:Eq:LotkaVolterra}. The inherent growth of species is given by $c_i x_i$. We chose $c_i=0.25\sum_{j}(a_{ji}-a_{ij})$. The diagonal of the  adjacency matrix is set to zero, $a_{ii}=0$, to prevent intraspecies competition. The initial population is uniformly distributed between 0 and 1. The integration is done using an eighth order Runge-Kutta method.

\subsubsection*{Noisy neural dynamics}
Neural dynamics on the C. Elegans are simulated using the noisy theta model. First, we integrate the ODEs for the the phase of oscillators,
\begin{align}
	\dot{y}_i = [1-&\cos(y_i)]\nonumber \\&+ [1+\cos(y_i)]\left[I_i+\theta N^{-1}\sum_{j=1}^N a_{ij}[1-\cos(y_j)]\right], \label{SI:Eq:thetamodel}
\end{align}
where $y_i(0)$ is initialized uniformly between 0 and $2\pi$. The parameters $I_i$ control the noise level and are sampled from a uniform distribution $U(0.05,0.15)$ and $\theta=5$. We integrate the system using an Euler integrator with $dt=0.1$. The external current $I_i$ is sampled on each time step. A spike is said to be produced when $\cos(y_i)>0.95$,  i.e., when $y_i$ is close to $\pi$. The continuous variable $y_i$ is transformed into a discrete state variable $x_i = H[\cos(y_i)-0.95]$ where $H(y)=1$ if $y>0$ and $H(y)=0$ otherwise. The time series $\bm{y}$ is then discarded and only $\bm{x}$ is used for inference.

\newcommand{\GINConv}{\text{GINConv}}
\newcommand{\Linear}{\bm{W}}
\subsubsection*{GNN model architecture} 
We have trained graph neural networks to forecast the activity. For a single forecast of the activity of node $i$, the GNN takes as input an adjacency matrix $\bm{A}$ and a window $\bm{X}^{(a)}=\lbrace \bm{x}_i^{(a)} \rbrace_{i=1...N}\in\mathbb{R}^{a\times N}$ of length $a$ of consecutive time steps of activity of all $N$ nodes, and it outputs the forecasted activity $\hat{\bm{x}}=\lbrace \hat{x}_i \rbrace_{i=1...N}$ of all nodes,
\begin{equation}
	\hat{\bm{x}} = \text{GNN}(\bm{A}, \bm{X}^{(a)}, \bm{\Lambda}).\label{EQ:GNN_single}
\end{equation}
The GNN is formed of two main operations. For an individual node $i$, 
\begin{subequations}
	\label{eq:gnn_decomposed}
	\begin{align}
		\bm{y}_i &= h_{\bm{\Lambda}} \left( \bm{x}^{(a)}_i + \sum_{j=1}^{N} a_{ij}\bm{x}^{(a)}_j\right)\label{C4:Eq:GinConv1}\\
			\hat{x}_i &= F(\bm{y}_i),
	\end{align}
\end{subequations}
where \eqref{C4:Eq:GinConv1} is a parametrized and non-linear operation called a GINConv layer \cite{xu2018powerful}, and $F(\cdot)$ is a trainable function. The operation $h_{\bm{\Lambda}}$ is a sequence of linear transformations and ReLU activations with the exact form detailed in the Supplementary information. 

For directed networks, \eqref{EQ:GNN_single} is replaced with a linear transformation of two distinct neural networks
\begin{equation}
	\hat{\bm{x}}= \Linear[\text{GNN}(\bm{A}, \bm{X}^{(a)}, \bm{\Lambda}_1)\| \text{GNN}(\bm{A}^T, \bm{X}^{(a)},\bm{\Lambda}_2)].
\end{equation}
All GNN model parameters $\bm{\Lambda}$ are randomly initialized using Kaiming's method \cite{he2015delving}.
\subsubsection*{GNN training} We first initialize $\pgnn$ so that $\sigma(\pgnn)=10^{-2}$ for existing edges of $\bm{A}$ and 0 otherwise. 

A training step goes as follows. We randomly slice a window at randomly chosen time of length $a+1$ in $\bm{X}$. The first $a$ values, $\bm{X}^{(a)}$, are given to the model while the last values are used as the forecasting target $\bm{x}$. We used $a=1$ for the SIS and predator-prey dynamics, and $a=20$ for the neural dynamics. The neural dynamics gets a longer activity history because the actual state of the system is hidden due to the binary spike transformation. Therefore, the information from the observed state is only partial and the time series are non-Markovian. For these reasons, the task is inherently more difficult and requires a longer activity history. 

We apply a forward step following the prescription of \eqref{C4:Eq:GNN}.
We optimize the target forecast over a cross-entropy loss for binary dynamics (epidemics, neural) and the absolute error loss for continuous dynamics (ecological) with the RAdam optimizer \cite{liu2019variance}. 

We use a learning rate of $10^{-3}$ for the parameters $\bm{\Lambda}$, with a decay of 0.1 after 1000 steps, and a learning rate of $10^{-2}$ for the perturbation matrix $\pgnn$. The model is optimized on $10^5$ steps. It roughly takes 30 seconds per training on a personal computer to achieve an inference over a network of $N=100$ nodes with time series of length $T=300$.

\newcommand{\bmABayes}{\bm{A}^*}
\subsubsection*{Bayesian model}
\newcommand{\pbayes}{\bm{p}}
\newcommand{\pp}{q}
We estimate the perturbed structure from the expected value of the posterior distribution as
\begin{equation}
	\hat{\bm{A}'}=\sum_{\bmABayes}\bmABayes\Pr(\bmABayes| \bm{X}, \bm{A}, \bm{\theta}),\label{C4:Eq:MCMC:posterior}
\end{equation}
and the perturbation matrix from $\hat{\bm{P}}=\bm{A}-\hat{\bm{A}'}$. The posterior can be rewritten using Bayes theorem
\begin{equation}
	\Pr(\bm{A}^*|\bm{X},\bm{A}, \bm{\theta}) \propto \Pr(\bm{X}|\bm{A},\bm{A}^*, \bm{\theta}) \Pr(\bm{A}^*|\bm{A}),
\end{equation}
where $\Pr(\bm{X}|\bm{A},\bm{A}^*, \bm{\theta})$ is the likelihood and $\Pr(\bm{A}^*|\bm{A})$ is the prior distribution, that we consider uniform.
Moreover, since only the portion $t > \tau$ is needed to infer the perturbed matrix, we can drop the dependence on $\bm{A}$ in the likelihood, i.e.,  $\Pr(\bm{X}|\bm{A}^*, \bm{\theta})$.

The posterior of \eqref{C4:Eq:MCMC:posterior} is sampled using a Metropolis-Hasting scheme \cite{metropolis1953equation}. We denote the adjacency matrix at sampling step $\nu$ by $\bm{A}_{\nu}$ that we have initialized with the unperturbed graph $\bm{A}_0=\bm{A}$.

At each sampling step, we randomly select an edge $(i,j) \in \mathcal{E}$. If the edge $(i,j)$ is present ($[\bm{A}_{\nu}]_{ij}=1$), we remove it, $[\bmABayes_{\nu+1}]_{ij}=0$, otherwise we add it back, $[\bmABayes_{\nu+1}]_{ij}=1$. Since graphs are undirected for the SIS dynamics, each graph proposition is symmetrically applied, e.g., $[\bmABayes_{\nu+1}]_{ij}=[\bmABayes_{\nu+1}]_{ji}$,.

The probability of accepting the proposition $\bm{A}_{\nu}\to\bmABayes_{\nu+1}$ at step $\nu+1$ is given by
\begin{equation}
	\rho = \min\left(1, \dfrac{\Pr(\bm{X}| \bmABayes_{\nu+1}, \bm{\theta})}{\Pr(\bm{X}|\bm{A}_{\nu}, \bm{\theta})}\right).
\end{equation}
The likelihood $\Pr(\bm{X}|\bm{A}^*, \bm{\theta})$ of the SIS model can be approximated by the product over individual transition probabilities for each node
\begin{align}
    \Pr(\bm{X}|\bm{A}^*, \bm{\theta}) = \prod_{i = 1}^N \prod_{t > \tau}^T \Pr[x_i(t) | x_i(t-1); \bm{A}^*,\bm{X}(t-1),\bm{\theta}] \;.\label{C4:Eq:sis:likelihood}
\end{align}
which is exact if the process is noiseless, $p_\mathrm{flip} \equiv \pp = 0$. For a node $i$ of degree $k$, with $l \equiv \sum_{j} a_{ij} x_j(t-1)$ infected neighbors, $x_i(t-1) \equiv v$ and $x_i(t) \equiv w$, the transition probabilities are
\begin{align}
    \mathrm{Pr}(w | v; l, k,\bm{\theta}) =& R(v,\pp)  R[w,R(\beta,\pp)] \;, \notag \\
    &+ [1- R(v,\pp)] R[w,R(F_{kl},\pp)] \;,
\end{align}
where $R(y,z) \equiv y(1-z) + (1-y)z$ and
\begin{align}
    F_{kl} = [\pp + (1-\pp)(1-\alpha)]^l[1- \pp + \pp(1-\alpha)]^{k-l} \;.
\end{align}
Recall that $\alpha$ and $\beta$ are given SIS parameters, and $\pp$ is the flip probability. Complete derivation is covered in the Supplementary information. We sample the posterior distribution over $\Gamma=2000$ propositions. We estimate the posterior distribution of \eqref{C4:Eq:MCMC:posterior} by counting the relative number of steps held by a certain graph, i.e., $\Pr(\bmABayes| \bm{X},\bm{A}, \bm{\theta})=\Gamma^{-1}\sum_{\nu} \mathbb{I}(\bm{A}_{\nu}-\bmABayes)$ where $\mathbb{I}(\cdot)=1$ only if the argument is element-wise zero and $\mathbb{I}(\cdot)=0$ otherwise.

\subsubsection*{Functional reconstruction methods}
\newcommand{\STD}{\text{std}}
The correlation matrix has elements
\begin{equation}
	\hat{a}_{ij}= \left|\dfrac{(\bm{x}_i-\bm{1}\mu_i)^T(\bm{x}_j-\bm{1}\mu_j)}{\sigma_i\sigma_j}\right|\label{C4:Eq:Correlation},
\end{equation}
where $\mu_i=\langle \bm{x}_i\rangle$ is the average activity of node $i$ and $\sigma_i$ is the standard deviation of $\bm{x}_i$. 
In the Granger causality method \cite{granger1969investigating}, the element $(i,j)$ of the adjacency matrix is estimated as the ratio of the standard deviation of the forecasting error of linear models $F(\cdot)$ trained with solely the activity of node $i$ and the other, $G(\cdot,\cdot)$, trained with the activity of nodes $i$ and $j$, 
\begin{equation}
	\hat{a}_{ij} = \dfrac{\STD[\bm{x}_i-F(\bm{x}_i)]}{\STD[\bm{x}_i-G(\bm{x}_i, \bm{x}_j)]}.
\end{equation}
For both methods, the estimated adjacency matrix $\hat{\bm{A}}'$ is computed using only the nodes activity after the perturbation. To suppress initially absent edges, we also perform an element-wise multiplication between $\hat{\bm{A}}'$ and the original adjacency matrix $\bm{A}$. If the network is undirected, we average the score over in and out edges, i.e., $\hat{\bm{A}}' \leftarrow \frac{1}{2}(\hat{\bm{A}}'+\hat{\bm{A}}'^T)$. We then normalize the estimated adjacency matrix and evaluate the perturbation matrix by $\hat{\bm{P}}= \bm{A}-\bm{\hat{A}}'$.

\newcommand{\TPR}{\text{TPR}}
\newcommand{\FPR}{\text{FPR}}
\subsubsection*{Evaluation metrics} Predicted removed edges are compared to ground-truth edges under a Receiver Operating Characteristic (ROC) curve analysis. For each inference, models output a weighted matrix of perturbation $\hat{\bm{P}}$. Note that only existing edges of $\bm{A}$ are assigned to a non-zero score because adding edges is prohibited. Perturbation scores are then rescaled between 0 and 1 where a score of $1$ is the most likely of being removed. Then, we compute the True/False Positive Rates (\TPR/\FPR) under a varying threshold $r$ from 0 to 1:
\begin{equation}
	\TPR(r) = \dfrac{N_{\text{TP}}(r)}{N_{\text{TP}(r)}+N_{\text{FN}(r)}}
\end{equation}
\begin{equation}
	\FPR(r) = \dfrac{N_{\text{FP}}(r)}{N_{\text{FP}(r)}+N_{\text{TN}}(r)},
\end{equation}
where the positive edges are those with a higher score than the threshold, i.e., the edge $(i,j)$ with $\hat{P}_{ij}>r$. The AUC is computed as the area under the ROC curve and is independent of the threshold $r$ applied. The \auc{} can be interpreted as the probability that the model distinguishes whether an edge is present or absent from the perturbation set $d\mathcal{E}$.

\subsubsection*{Acknowledgments}
This work was funded by the Fonds de recherche du Québec-Nature et technologies, the Natural Sciences and Engineering Research Council of Canada, and Sentinel North, financed by the Canada First Research Excellence Fund. The authors want to thank Antoine Allard, Patrick Desrosiers, and Louis J. Dubé for supports and helpful comments, Mohamed Babine for fruitful discussions and code sharing, and Martin Laprise for useful comments and for allocating computational resources. We acknowledge Calcul Québec for using their computing facilities.
\end{spacing}
\bibliographystyle{siam}
\bibliography{refs}

\begin{thebibliography}{10}

\bibitem{angelini2011ecosystem}
{\sc R.~Angelini and F.~Vaz-Velho}, {\em Ecosystem structure and trophic
  analysis of {A}ngolan fishery landings}, Sci. Mar., 75 (2011), pp.~309--319.

\bibitem{barabasi2009scale}
{\sc A.-L. Barab{\'a}si}, {\em Scale-free networks: {A} decade and beyond},
  Science, 325 (2009), pp.~412--413.

\bibitem{barabasi1999emergence}
{\sc A.-L. Barab{\'a}si and R.~Albert}, {\em Emergence of scaling in random
  networks}, Science, 286 (1999), pp.~509--512.

\bibitem{barzel2013network}
{\sc B.~Barzel and A.-L. Barab{\'a}si}, {\em Network link prediction by global
  silencing of indirect correlations}, Nat. Biotechnol., 31 (2013),
  pp.~720--725.

\bibitem{boerlijst2013catastrophic}
{\sc M.~C. Boerlijst, T.~Oudman, and A.~M. de~Roos}, {\em Catastrophic collapse
  can occur without early warning: {E}xamples of silent catastrophes in
  structured ecological models}, PloS one, 8 (2013), p.~e62033.

\bibitem{bressler2011wiener}
{\sc S.~L. Bressler and A.~K. Seth}, {\em Wiener--{G}ranger causality: a well
  established methodology}, Neuro{I}mage, 58 (2011), pp.~323--329.

\bibitem{brugere2018network}
{\sc I.~Brugere, B.~Gallagher, and T.~Y. Berger-Wolf}, {\em Network structure
  inference, a survey: {H}otivations, methods, and applications}, ACM Computing
  Surveys (CSUR), 51 (2018), pp.~1--39.

\bibitem{chen2006frequency}
{\sc Y.~Chen, S.~L. Bressler, and M.~Ding}, {\em Frequency decomposition of
  conditional {G}ranger causality and application to multivariate neural field
  potential data}, J. Neurosci. Methods, 150 (2006), pp.~228--237.

\bibitem{clements2019early}
{\sc C.~F. Clements, M.~A. McCarthy, and J.~L. Blanchard}, {\em Early warning
  signals of recovery in complex systems}, Nat. Commun., 10 (2019), p.~1681.

\bibitem{dakos2015resilience}
{\sc V.~Dakos, S.~R. Carpenter, E.~H. van Nes, and M.~Scheffer}, {\em
  Resilience indicators: prospects and limitations for early warnings of regime
  shifts}, Philos. Trans. R. Soc. B, 370 (2015), p.~20130263.

\bibitem{ermentrout1986parabolic}
{\sc G.~B. Ermentrout and N.~Kopell}, {\em Parabolic bursting in an excitable
  system coupled with a slow oscillation}, SIAM Journal on Applied Mathematics,
  46 (1986), pp.~233--253.

\bibitem{feizi2013network}
{\sc S.~Feizi, D.~Marbach, M.~M{\'e}dard, and M.~Kellis}, {\em Network
  deconvolution as a general method to distinguish direct dependencies in
  networks}, Nat. Biotechnol., 31 (2013), p.~726.

\bibitem{gao2016universal}
{\sc J.~Gao, B.~Barzel, and A.-L. Barab{\'a}si}, {\em Universal resilience
  patterns in complex networks}, Nature, 530 (2016), pp.~307--312.

\bibitem{granger1969investigating}
{\sc C.~W. Granger}, {\em Investigating causal relations by econometric models
  and cross-spectral methods}, Econometrica,  (1969), pp.~424--438.

\bibitem{hamilton2017inductive}
{\sc W.~Hamilton, Z.~Ying, and J.~Leskovec}, {\em Inductive representation
  learning on large graphs}, in NeurIPS, 2017, pp.~1024--1034.

\bibitem{he2015delving}
{\sc K.~He, X.~Zhang, S.~Ren, and J.~Sun}, {\em Delving deep into rectifiers:
  {S}urpassing human-level performance on {I}magenet classification}, in
  Proceedings of the ICCV, 2015, pp.~1026--1034.

\bibitem{hecker2009gene}
{\sc M.~Hecker, S.~Lambeck, S.~Toepfer, E.~Van~Someren, and R.~Guthke}, {\em
  Gene regulatory network inference: data integration in dynamic models—a
  review}, Biosystems, 96 (2009), pp.~86--103.

\bibitem{hoegh2007coral}
{\sc O.~Hoegh-Guldberg, P.~J. Mumby, A.~J. Hooten, R.~S. Steneck,
  P.~Greenfield, E.~Gomez, C.~D. Harvell, P.~F. Sale, A.~J. Edwards,
  K.~Caldeira, et~al.}, {\em Coral reefs under rapid climate change and ocean
  acidification}, Science, 318 (2007), pp.~1737--1742.

\bibitem{jager2019systematically}
{\sc G.~J{\"a}ger and M.~F{\"u}llsack}, {\em Systematically false positives in
  early warning signal analysis}, PloS one, 14 (2019), p.~e0211072.

\bibitem{jiang2018predicting}
{\sc J.~Jiang, Z.-G. Huang, T.~P. Seager, W.~Lin, C.~Grebogi, A.~Hastings, and
  Y.-C. Lai}, {\em Predicting tipping points in mutualistic networks through
  dimension reduction}, Proc. Natl. Acad. Sci. U.S.A., 115 (2018),
  pp.~E639--E647.

\bibitem{kaiser2006nonoptimal}
{\sc M.~Kaiser and C.~C. Hilgetag}, {\em Nonoptimal component placement, but
  short processing paths, due to long-distance projections in neural systems},
  PLOS Comput. Biol., 2 (2006).

\bibitem{kefi2013early}
{\sc S.~K{\'e}fi, V.~Dakos, M.~Scheffer, E.~H. Van~Nes, and M.~Rietkerk}, {\em
  Early warning signals also precede non-catastrophic transitions}, Oikos, 122
  (2013), pp.~641--648.

\bibitem{kipf2016semi}
{\sc T.~N. Kipf and M.~Welling}, {\em Semi-supervised classification with graph
  convolutional networks}, arXiv preprint arXiv:1609.02907,  (2016).

\bibitem{laurence2019spectral}
{\sc E.~Laurence, N.~Doyon, L.~J. Dub{\'e}, and P.~Desrosiers}, {\em Spectral
  dimension reduction of complex dynamical networks}, Phys. Rev. X, 9 (2019),
  p.~011042.

\bibitem{lenton2011early}
{\sc T.~M. Lenton}, {\em Early warning of climate tipping points}, Nat. Clim.
  Change, 1 (2011), pp.~201--209.

\bibitem{liu2019variance}
{\sc L.~Liu, H.~Jiang, P.~He, W.~Chen, X.~Liu, J.~Gao, and J.~Han}, {\em On the
  variance of the adaptive learning rate and beyond}, arXiv preprint
  arXiv:1908.03265,  (2019).

\bibitem{lynall2010functional}
{\sc M.-E. Lynall, D.~S. Bassett, R.~Kerwin, P.~J. McKenna, M.~Kitzbichler,
  U.~Muller, and E.~Bullmore}, {\em Functional connectivity and brain networks
  in schizophrenia}, Journal of Neuroscience, 30 (2010), pp.~9477--9487.

\bibitem{mann1996robust}
{\sc M.~E. Mann and J.~M. Lees}, {\em Robust estimation of background noise and
  signal detection in climatic time series}, Clim. Change, 33 (1996),
  pp.~409--445.

\bibitem{marbach2012wisdom}
{\sc D.~Marbach, J.~C. Costello, R.~K{\"u}ffner, N.~M. Vega, R.~J. Prill, D.~M.
  Camacho, K.~R. Allison, A.~Aderhold, R.~Bonneau, Y.~Chen, et~al.}, {\em
  Wisdom of crowds for robust gene network inference}, Nat. Methods, 9 (2012),
  p.~796.

\bibitem{may2008ecology}
{\sc R.~M. May, S.~A. Levin, and G.~Sugihara}, {\em Ecology for bankers},
  Nature, 451 (2008), pp.~893--894.

\bibitem{metropolis1953equation}
{\sc N.~Metropolis, A.~W. Rosenbluth, M.~N. Rosenbluth, A.~H. Teller, and
  E.~Teller}, {\em Equation of state calculations by fast computing machines},
  J. Chem. Phys., 21 (1953), pp.~1087--1092.

\bibitem{newman2018estimating}
{\sc M.~E. Newman}, {\em Estimating network structure from unreliable
  measurements}, Phys. Rev. E, 98 (2018), p.~062321.

\bibitem{newman2018networks}
\leavevmode\vrule height 2pt depth -1.6pt width 23pt, {\em Networks}, Oxford
  university press, 2018.

\bibitem{pan2016predicting}
{\sc L.~Pan, T.~Zhou, L.~L{\"u}, and C.-K. Hu}, {\em Predicting missing links
  and identifying spurious links via likelihood analysis}, Sci. Rep., 6 (2016),
  pp.~1--10.

\bibitem{park2013structural}
{\sc H.-J. Park and K.~Friston}, {\em Structural and functional brain networks:
  {F}rom connections to cognition}, Science, 342 (2013), p.~1238411.

\bibitem{pastor2015epidemic}
{\sc R.~Pastor-Satorras, C.~Castellano, P.~Van~Mieghem, and A.~Vespignani},
  {\em Epidemic processes in complex networks}, Rev. Mod. Phys., 87 (2015),
  p.~925.

\bibitem{peixoto2019network}
{\sc T.~P. Peixoto}, {\em Network reconstruction and community detection from
  dynamics}, Phys. Rev. Lett., 123 (2019), p.~128301.

\bibitem{perez2009pressure}
{\sc R.~P{\'e}rez, V.~Puig, J.~Pascual, A.~Peralta, E.~Landeros, and
  L.~Jordanas}, {\em Pressure sensor distribution for leak detection in
  {B}arcelona water distribution network}, Water Sci. Tech.-W. Sup., 9 (2009),
  pp.~715--721.

\bibitem{pires2017rewilding}
{\sc M.~M. Pires}, {\em Rewilding ecological communities and rewiring
  ecological networks}, Perspect. Ecol. Conser., 15 (2017), pp.~257--265.

\bibitem{roebroeck2005mapping}
{\sc A.~Roebroeck, E.~Formisano, and R.~Goebel}, {\em Mapping directed
  influence over the brain using {G}ranger causality and {fMRI}}, Neuro{I}mage,
  25 (2005), pp.~230--242.

\bibitem{scheffer2010foreseeing}
{\sc M.~Scheffer}, {\em Foreseeing tipping points}, Nature, 467 (2010),
  pp.~411--412.

\bibitem{scheffer2009early}
{\sc M.~Scheffer, J.~Bascompte, W.~A. Brock, V.~Brovkin, S.~R. Carpenter,
  V.~Dakos, H.~Held, E.~H. Van~Nes, M.~Rietkerk, and G.~Sugihara}, {\em
  Early-warning signals for critical transitions}, Nature, 461 (2009), p.~53.

\bibitem{schindler2006recent}
{\sc D.~W. Schindler}, {\em Recent advances in the understanding and management
  of eutrophication}, Limnol. Oceanogr., 51 (2006), pp.~356--363.

\bibitem{seth2015granger}
{\sc A.~K. Seth, A.~B. Barrett, and L.~Barnett}, {\em Granger causality
  analysis in neuroscience and neuroimaging}, J. Neurosci., 35 (2015),
  pp.~3293--3297.

\bibitem{shandilya2011inferring}
{\sc S.~G. Shandilya and M.~Timme}, {\em Inferring network topology from
  complex dynamics}, New J. Phys., 13 (2011), p.~013004.

\bibitem{sheikhattar2018extracting}
{\sc A.~Sheikhattar, S.~Miran, J.~Liu, J.~B. Fritz, S.~A. Shamma, P.~O. Kanold,
  and B.~Babadi}, {\em Extracting neuronal functional network dynamics via
  adaptive {G}ranger causality analysis}, Proc. Natl. Acad. Sci. U.S.A., 115
  (2018), pp.~E3869--E3878.

\bibitem{wilkat2019no}
{\sc T.~Wilkat, T.~Rings, and K.~Lehnertz}, {\em No evidence for critical
  slowing down prior to human epileptic seizures}, Chaos, 29 (2019), p.~091104.

\bibitem{wissel1984universal}
{\sc C.~Wissel}, {\em A universal law of the characteristic return time near
  thresholds}, Oecologia, 65 (1984), pp.~101--107.

\bibitem{xu2018powerful}
{\sc K.~Xu, W.~Hu, J.~Leskovec, and S.~Jegelka}, {\em How powerful are graph
  neural networks?}, arXiv preprint arXiv:1810.00826,  (2018).

\bibitem{young2019reconstruction}
{\sc J.-G. Young, F.~S. Valdovinos, and M.~E. Newman}, {\em Reconstruction of
  plant--pollinator networks from observational data}, bio{R}xiv,  (2019),
  p.~754077.

\end{thebibliography}


\begin{thebibliography}{1}

\bibitem{hamilton2017inductive}
{\sc W.~Hamilton, Z.~Ying, and J.~Leskovec}, {\em Inductive representation
  learning on large graphs}, in NeurIPS, 2017, pp.~1024--1034.

\bibitem{newman2018networks}
{\sc M.~E. Newman}, {\em Networks}, Oxford university press, 2018.

\bibitem{velivckovic2017graph}
{\sc P.~Veli{\v{c}}kovi{\'c}, G.~Cucurull, A.~Casanova, A.~Romero, P.~Lio, and
  Y.~Bengio}, {\em Graph attention networks}, arXiv preprint arXiv:1710.10903,
  (2017).

\bibitem{xu2018powerful}
{\sc K.~Xu, W.~Hu, J.~Leskovec, and S.~Jegelka}, {\em How powerful are graph
  neural networks?}, arXiv preprint arXiv:1810.00826,  (2018).

\end{thebibliography}

\end{document}


\maketitle

{\hypersetup{linkcolor=black}
\tableofcontents
}

\section{GNN Model} 
\subsection{Layers}
The input of the GNN model for the target node $i$ is its current activity $\bm{x}_i^{(a)}$, and the activity of its neighbors $\bm{x}_j^{(a)}$ for all $j$ with $a_{ij}=1$. Here, $\bm{x}_j^{(a)}$ denotes a slice of $a$ consecutive time steps of node $j$ activity.

The length $a$ is a parameter that must be set beforehand. In practice, it is best if set to the expected memory of the dynamical mechanism. For instance, the memory of a markovian process is $a=1$. If $a$ is too small, then the input activity will not be sufficient to forecast the future activity. On the contrary, if $a$ is much larger than the memory of the dynamical system, the model should learn to vanish the weights allocated for the oldest activity steps. A large $a$ also exposes the model to overfitting by finding superfluous patterns in previous states of activity.

For a single forward pass, the GNN model applies the following operations:
\begin{subequations}
	\label{Eq:GNNModel:FG}
	\begin{align}
		\bm{y}_i &= h_{\bm{\Lambda}} \left( \bm{x}_i^{(a)} + \sum_{j=1}^{N} a_{ij}\bm{x}_j^{(a)}\right)\label{Eq:GNNModel:FGa}\\
		\hat{x}_i &= F(\bm{y}_i)\label{Eq:GNNModel:FGb}
	\end{align}
\end{subequations}
where $h_{\bm{\Lambda}}(\cdot)$ is a trainable layer, and $F(\cdot)$ is a function that may have trainable parameters. The function $h_{\bm{\Lambda}}(\cdot)$ is called a GINConv~\cite{xu2018powerful}. It aggregates the neighborhood of node $i$ by summing its neighbors' activity, and then by applying a non-linear operation described by $h_{\bm{\Lambda}}(\cdot)$. Thanks to the sum operation on $j$, the layer is independent of the number of neighbors of node $i$. It makes the GNN model inductive as neither the size of the graph, nor the number of neighbors, are hard-coded in the layers. 

During training, a loss function is computed between the model output and the ground-truth future activity,
\begin{equation}
	L(\hat{x}_i, x_i).
\end{equation}
The spreading and the neural dynamics use the binary cross entropy,
\begin{equation}
	L(\hat{x}_i, x_i) = x_i\log(\hat{x}_i)+(1-x_i)\log(1-\hat{x}_i),
\end{equation}
while the predator-prey dynamics uses the L1 loss,
\begin{equation}
	L(\hat{x}_i, x_i) = |x_i-\hat{x}_i|.
\end{equation}

The GINConv layer is the core operation of the GNN model. By explicitly using the adjacency matrix $\bm{A}$ in Eq.~\eqref{Eq:GNNModel:FGa}, the model is able to backpropagate the error into the adjacency matrix $\bm{A}$ which will be crucial for inferring the perturbation. Note that the backpropagation requires the mathematical operations applied by the layers to be differentiable, which obviously is satisfied when using the sum aggregation.

The GINConv could have been equivalently written as a sparse operation on the node's neighborhood $\mathcal{N}$,
\begin{equation}
	\bm{y}_i = h_{\bm{\Lambda}} \left( \bm{x}_i + \sum_{j\in \mathcal{N}_i} \bm{x}_j\right),
\end{equation}
where the independence with the number of nodes $N$ in the graph still holds. Yet, slicing beforehand the nodes activity directly on the neighborhood of the node $i$ would disable the differentiability of the adjacency matrix $\bm{A}$. Hence, the essential backpropagation of $\bm{A}$ would fail.\\

Table~\ref{Tb:GNNModel} summarizes the mathematical operations of the model for the different dynamics. There is no guarantee that this selection of hyperparameters, e.g., number of linear transformations or the number of hidden dimensions, is the best possible configuration for our task neither that it will fit to a different dynamics. From our experiments, similar results were achieved with other sets of hyperparameters. Worse performance resulted from a bad choice of learning rates rather than an inappropriate model design. An ablation study could be made in the future to validate our observations.

It is worth noting that other GNN layers could have been used instead of the GINConv, such as the Graph Attention Layer (GAT) \cite{velivckovic2017graph} and the SAGEConv \cite{hamilton2017inductive}.

We have used the GINConv mostly by convenience. Most deep learning frameworks of GNN models use a sparse representation of the adjacency matrix, which makes the adjacency matrix not differentiable. Only few layers have been coded by the PyTorch community for using a dense adjacency matrix. 

The function $F(\cdot)$ is an activation layer used to control the shape of the output and its bounds. It receives as input $\bm{y}_i$, i.e., the output of the GINConv. We list the functions used for the selected dynamics,
\begin{align}
	\text{SIS:}~~~&F(\bm{y}) = \dfrac{1}{1+e^{-\bm{y}}};\\
	\text{Predator-prey:}~~~&F(\bm{y}) = \bm{W}\bm{y}+\bm{c};\\
	\text{Neural dynamics:}~~~&F(\bm{y}) = \dfrac{1}{1+e^{-\bm{W}\bm{y}+\bm{c}}},
\end{align}
with detailed dimensions in Table~\ref{Tb:GNNModel}.

For the binary dynamics (SIS and neural dynamics), we use the sigmoid function to bound the output between 0 and 1. In doing so, the output can be interpreted as a probability that the next activity is the active state. 
For the predator-prey dynamics, the output is not required to be bounded. Therefore, we simply apply a linear transformation to match the desired output size. Note that $\bm{W}$ and $\bm{c}$ are trainable parameters.

\begin{table}
	\caption{Layer by layer description of the GNN models for each dynamics. For each sequence, the operations are applied in order from top to bottom. A linear operation Linear($n,m$) is defined as $f(\bm{x})=\bm{W}(\bm{x})+\bm{c}$ with $\bm{W}\in\mathbb{R}^{m\times n}$ and $\bm{c}\in\mathbb{R}^{m}$. The ReLU operation is the element-wise maximum $\text{ReLU}(\bm{x})=\text{max}(0,\bm{x})$. }
	\label{Tb:GNNModel}
	\centering
	\begin{tabular}{L{3cm}C{4cm}C{4cm}C{2cm}}
		\hline \\
		Dynamics & $h_{\bm{\Lambda}}(\cdot)$ & $F(\cdot)$ & Number of parameters \\\\
		\hline 
		\multirow{4}{*}{SIS} & Linear(1, 16) & \multirow{4}{*}{Sigmoid} & \multirow{4}{*}{1\,185} \\
		                      & ReLU & \\
		                      & Linear(16, 64) & \\
		                      & ReLU & \\
		                      & Linear(64, 1) & \\
		\hline 
		\multirow{4}{*}{Predator-prey} & Linear(1, 64) & \multirow{4}{*}{Linear(128, 1)$^*$} & \multirow{4}{*}{37\,569}\\
		                      & ReLU & \\
		                      & Linear(64, 128) & \\
		                      & ReLU & \\
		                      & Linear(128, 64) & \\
		\hline 
		\multirow{4}{*}{Neural} & Linear(20, 64) & & \multirow{4}{*}{17\,441}\\
		                      & ReLU &  Linear(64, 1)$^*$ \\
		                      & Linear(64, 64) & Sigmoid\\
		                      & ReLU & \\
		                      & Linear(64, 32) & \\

		\hline \\
		\multicolumn{3}{L{13cm}}{$^*${\small The linear function is twice the size of the $h_{\bm{\Lambda}}(\cdot)$ because the networks are directed and the outputs of two GINConvs are concatenated.}}\\
	\end{tabular}
\end{table}

For directed networks, we use two GINConvs so that the GNN model can distinguish between in- and out- edges,
\begin{subequations}
	\label{Eq:GNNModel:FG:directed}
	\begin{align}
		\bm{y}_i^{(1)} &= h^{(1)}_{\bm{\Lambda}} \left( \bm{x}_i + \sum_{j=1}^{N} a_{ij}\bm{x}_j\right)\label{Eq:GNNModel:FG:directed:a1}\\
		\bm{y}_i^{(2)} &= h^{(2)}_{\bm{\Lambda}} \left( \bm{x}_i + \sum_{j=1}^{N} a_{ji}\bm{x}_j\right)\label{Eq:GNNModel:FG:directed:a2}\\
		\hat{\bm{x}}_i &= F(\bm{y}_i^{(1)}\|\bm{y}_i^{(2)})\label{Eq:GNNModel:FG:directed:b}
	\end{align}
\end{subequations}
where $\cdot\|\cdot$ is the concatenation operator. This technique is essential to correctly infer on directed networks.  

\subsection{Weight decay}
L2 weight decay is often used to prevent overfitting. In our task, we found that using weight decay impairs the learning of the perturbation. When applying a weight decay on the trainable perturbation matrix $\pgnn$, we penalize the model on the norm of $|\pgnn|^2$. Therefore, it pushes the elements to zero, $[\pgnn]_{ij}\to 0$. Recall that the forecast is done using
\begin{equation}
	\hat{\bm{x}} = \text{GNN}(\bm{A}-\sigma(\pgnn), \bm{x}, \bm{\Lambda}).
\end{equation}
Hence, the direction $[\pgnn]_{ij}\to 0$ contributes to false positive detection $[\sigma(\pgnn)]_{ij}\to 0.5$. We would rather look for the direction $[\sigma(\pgnn)]_{ij}\to 0$ to push most edges toward a vanishing contributions except for truly removed edges. For these reasons, we did not use a weight decay in our simulations.

Note that the weight decay could be applied only on the inner parameters $\bm{\Lambda}$ without harming the perturbation. From experiments, using L2 weight decay did not lead to any noticeable improvements.

\section{Large perturbations and loopy scale-free networks}
\begin{figure}
	\centering
	\includegraphics[width=0.5\textwidth]{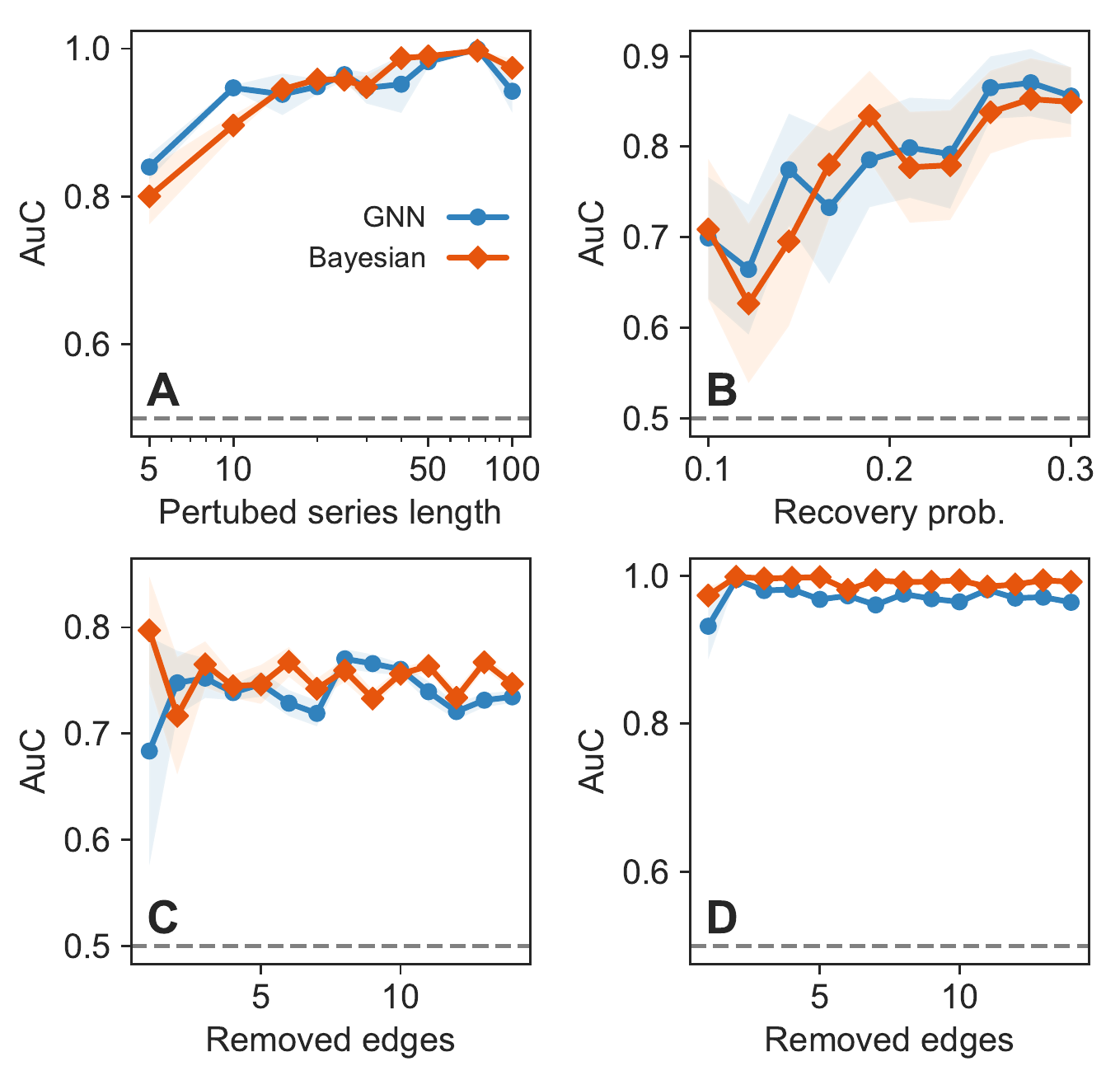}
	\caption{AUC on various schemes with SIS dynamics. In (A), we have varied when the change point $\tau$ occurs. The length $T-\tau$ of the perturbed time series is displayed on the x-axis (Scale-free graphs $N=100,\,m=1$ with $T=200$ and $\alpha=0.2,\beta=0.1$). (B) AUC on a scale-free graph with loops $N=100,m=3$ and $T=200,\alpha=0.1$. (C-D) AUC as a function of the number of removed edges (C: random graphs with $N=100,p=0.1, \alpha=0.1, \beta=0.1$) (D: scale-free graphs $N=100,m=1,\alpha=0.2,\beta=0.1$)}
	\label{Fig:SI:other_results}
\end{figure}

We explore how the model is affected by removing a large number of edges. From Fig.~\ref{Fig:SI:other_results}(C,D), we conclude that the GNN model performs almost as well as the Bayesian model in multiple edges perturbations. A slight deviation between the GNN and Bayesian model is noticeable for scale-free networks. This issue is detailed in the next section. Moreover, the performance measured with the AUC is quite stable up to 15 removed edges. In the considered scale-free networks of Fig.~\ref{Fig:SI:other_results}(D), the perturbation represents 15\% of all edges.  In Fig.~\ref{Fig:SI:other_results}(B), we present the AUC for loopy scale-free graphs. We conclude againt that the GNN model performs as well as the Bayesian approach.

\section{Challenging edges in scale-free networks}
\begin{figure}
	\centering
	\includegraphics[width=\textwidth]{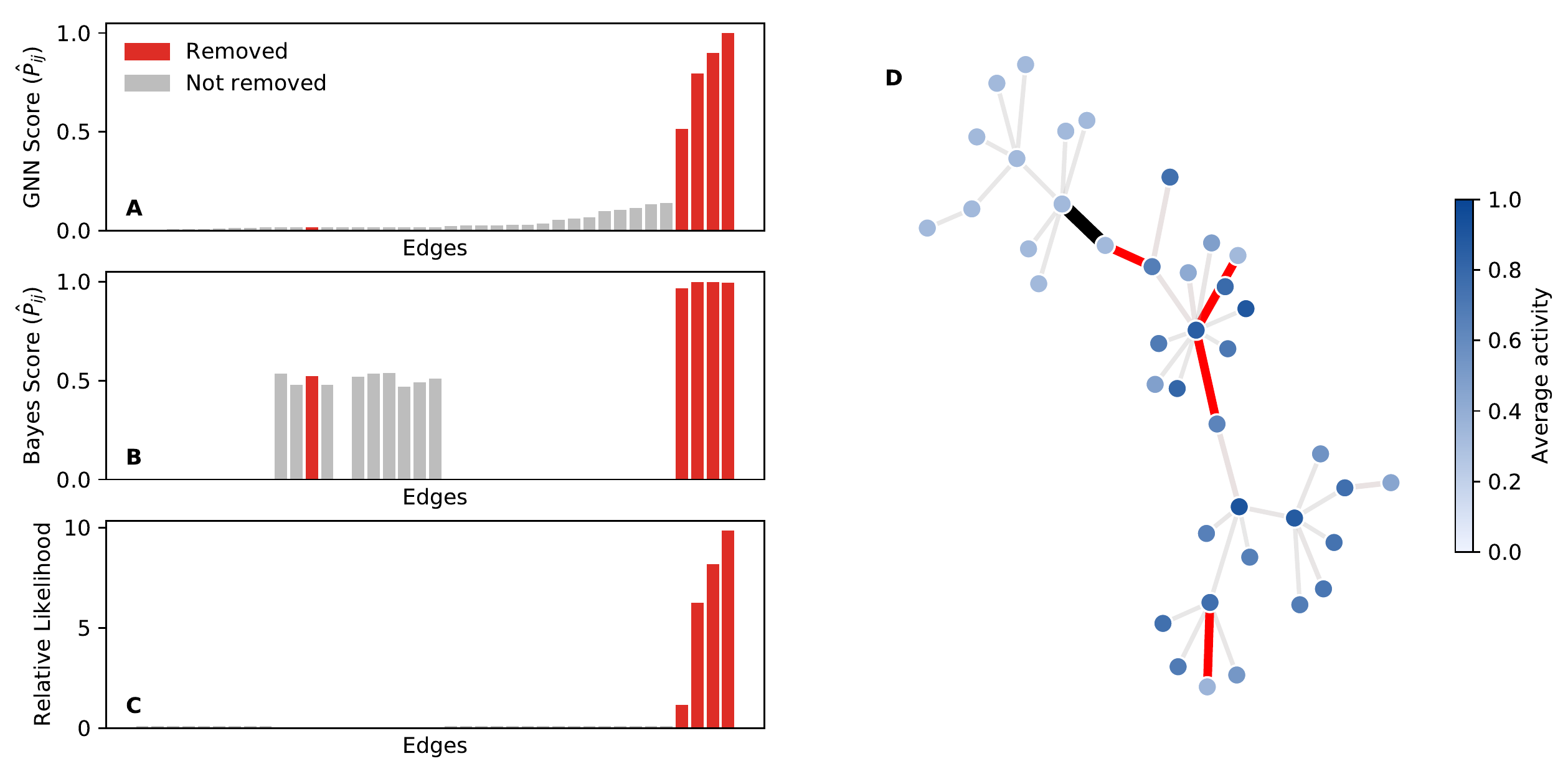}
	\caption{Example of inference on the SIS dynamics with $\alpha=0.12$ and $\beta=0.1$ over a scale-free networks with 5 removed edges. In (A-B), the scores of the infered perturbation matrix for all edges ordered by the GNN predicted scores. In (C), the relative variation of the likelihood $1-\Pr(\bm{X}|\bm{A}^*, \bm{\theta})/\Pr(\bm{X}|\bm{A}, \bm{\theta})$ when removing the edge ordered similarly as (A-B). In D, the network used for the example where red edges and the black edge are the perturbation. The black edge is a false negative according to the model. Only the red edges have been correctly predicted by the GNN model with a high score. The nodes are colored according to their average activity after the perturbation. }
	\label{Fig:SI:Example:bara:multiple}
\end{figure}

Interestingly, for scale-free networks, we observe that some removed edges can be more challenging to infer than usual. This emergent aspect is caused by removing core edges that disconnect the network into small inactive components. Therefore, if certain edges have been removed deeper down into the smaller components of the disconnected tree structure, they would be practically impossible to infer because of the absence of activity. The only time frame to infer them would be just after the perturbation, while the activity is rapidly decaying in the periphery. We show an example in Fig.~\ref{Fig:SI:Example:bara:multiple} where an edge from the perturbation does not induce a large likelihood drop. It is clearly shown that this edge connects two nodes with vanishing level of activity after the perturbation, which explains how difficult it can be to infer it.
 
\section{Length of the perturbed time series}
We investigate the effects of decreasing the size of the perturbed time series. More precisely, the moment of perturbation $\tau$ is moved closer to the end of the time series. In Fig.~\ref{Fig:SI:other_results}(A), we have simulated the dynamics over a fixed $T=200$ time steps. We execute the perturbation at various $t=\tau\in[0,200]$. We observe that the maximum AUC is achieved with only 10 perturbed time steps.

\newcommand{\pflip}{p_{\text{flip}}}
\section{Likelihood of the SIS model}

We now detail the likelihood of the SIS model. 
Consider the transition probability $\Pr[x_i(t) | x_i(t-1); \bm{A},\bm{X}(t-1),\bm{\theta}]$ for the noiseless SIS model $\pflip=0$. 

If the node $i$ is infected at time $t-1$, i.e., $x_i(t-1)=1$, then it may have recovered with probability $\beta$, i.e., $x_i(t)=0$,  or stayed infected with probability $1-\beta$, i.e., $x_i(t)=1$. Its contribution to the transition probability is
\begin{equation}
	x_i(t-1)[x_i(t)[1-\beta]+[1-x_i(t)]\beta].\label{SI:SIS:inactive}
\end{equation}
If the node is susceptible at time $t-1$, i.e., $x_i(t-1)=0$, it may get infected with probability $1-(1-\alpha)^{l_i}$, where ${l_i}=\sum_j a_{ij}x_j(t-1)$ is the number of infected neighbors, or stay susceptible with probability $(1-\alpha)^{l_i}$. Its contribution to the transition probability is
\begin{equation}
	[1-x_i(t-1)][x_i(t)[1-(1-\alpha)^{l_i}]+[1-x_i(t)](1-\alpha)^{l_i}].\label{SI:SIS:active}
\end{equation}
Combining \eqref{SI:SIS:inactive} and \eqref{SI:SIS:active}, the transition probabilities are
\begin{subequations}
	\label{SI:SIS:eqs:transition}
	\begin{equation}
	 	\Pr[w|v; \bm{A},\bm{X}(t-1),\alpha, \beta] = vR(w, \beta)+ [1-v]R(w, g_l)
	\end{equation} 
	where $w\equiv x_i(t)$, $v\equiv x_i(t-1)$, $g_l=(1-\alpha)^{l_i}$, and
	\begin{equation}
		R(y,z) = y(1-z) + (1-y)z. 
	\end{equation}
\end{subequations}

Because this is a Markov process, the likelihood of a time series is the product of the transition probabilities, 
\begin{align}
    \Pr(\bm{X}|\bm{A}, \bm{\theta}) = \prod_{i = 1}^N \prod_{t > \tau}^T \Pr[x_i(t) | x_i(t-1); \bm{A},\bm{X}(t-1),\bm{\theta}] \;.\label{Eq:sis:likelihood}
\end{align}

We now consider the SIS model with a flip probability $\pflip$. We introduce the observed variables $\bm{X}$ and the noiseless activity $\bm{Y}$, i.e., equivalent to the true activity before executing the flips. Note that since we do not observe the whole state of the system ($\bm{X}$ and $\bm{Y}$), correlation remains between $\bm{X}(t)$ and $\bm{X}(t-2)$, even after conditioning on $\bm{X}(t-1)$. Formally speaking, the process on the variables $\bm{X}$ alone is non-Markovian. However, we expect the correlation to decrease quickly with time, so we use a Markovian approximation to write the likelihood. In other words, we use only $\bm{X}(t-1)$ to describe $\bm{X}(t)$, as before.

First, consider that the node $i$ is truly infected at time $t-1$, i.e., $y_i(t-1)=1$. Assuming that we observe $x_i(t-1)$, we can write the conditional probability by considering that the flip succeeded with probability $\pflip$ and failed with probability $1-\pflip$,
\begin{equation}
	\Pr(y_i(t-1)=1|v) = v[1-\pflip] + [1-v]\pflip \equiv R[v,\pflip],
\end{equation}
where $v=x_i(t-1)$.
For a truly infected node, two processes take place. First, the node can recover with probability $\beta$ or stay infected with probability $1-\beta$. Second, the flip process succeeds with probability $\pflip$ or fails with probability $1-\pflip$. This composition is enclosed in $R[x_i(t), R(\beta, \pflip)]$. Therefore, the first contribution to the transition probability is
\begin{equation}
	R[x_i(t-1),\pflip] R[x_i(t), R(\beta, \pflip)].\label{SI:EQ:SIS:ACTIFLIP2}
\end{equation}
The second contribution to the transition probability adopt that the node $i$ is truly susceptible at time $t-1$, i.e., $y_i(t-1)=0$. This occurs with probability $1-R[x_i(t-1),\pflip]$. Again, two processes form the step: The possible infection of node $i$ and the flip process with probability $\pflip$. Therefore, the second contribution is
\begin{equation}
	(1-R[x_i(t-1),\pflip]) R[x_i(t), R(F_{i}, \pflip)].\label{SI:EQ:SIS:ACTIFLIP}
\end{equation}
where $F_{i}$ is the probability that the node $i$ \textit{does not get infected by its neighbors}, i.e., $\Pr(y_i(t)=0| y_i(t-1)=0, \bm{A}, \alpha, \beta)$. To obtain $F_{i}$, we need to consider the complete neighborhood of the node $i$. If a neighbor $j$ is observed infected $x_j(t-1)=1$, then two events could have occured:
\begin{enumerate}
	\item Node $j$ is infected at time $t-1$, i.e., $y_j(t-1)=1$, in which case the flip failed with probability $(1-\pflip)$. It fails to infect node $i$ with probability $1-\alpha$;
	\item Node $j$ is susceptible at time $t-1$, i.e, $y_j(t-1)=0$, in which case the flip noise succeed with probability $\pflip$, and it failed to infect node $i$ with probability $1$.
\end{enumerate}
As we observe $l_i$ infected nodes, the contribution of observed infected neighbors is
\begin{equation}
	[\pflip+(1-\pflip)(1-\alpha)]^{l_i}.\label{SI:Eq:SIS:infectedasdasd}
\end{equation}
For a node $i$ with $k_i$ neighbors, a similar argument can be made for the $k_i-l_i$ susceptible neighbors of node $i$, i.e., $x_j(t-1)=0$ for all $j$ that $a_{ij}=1$. The total contribution is
\begin{equation}
	[(1-\pflip)+\pflip(1-\alpha)]^{k_i-l_i}.\label{SI:Eq:SIS:infectedasdasd2}
\end{equation}
Combining \eqref{SI:Eq:SIS:infectedasdasd} with \eqref{SI:Eq:SIS:infectedasdasd2}, we obtain
\begin{equation}
	F_{i} = [\pflip + (1-\pflip)(1-\alpha)]^{l_i}[1- \pflip + \pflip(1-\alpha)]^{k_i-l_i}.\label{SI:EQ:FKL}
\end{equation}
From \eqref{SI:EQ:SIS:ACTIFLIP2} and \eqref{SI:EQ:SIS:ACTIFLIP}, the transition probabilities of the SIS model with flips are
\begin{equation}
 	\Pr[w|v; \bm{A},\bm{X}(t-1),\alpha, \beta] =R[v,\pflip] R[w, R(\beta, \pflip)]+(1-R[v,\pflip]) R[w, R(F_{i}, \pflip)]\label{SI:SIS:eqs:transition2}
\end{equation} 
where $w\equiv x_i(t)$, $v\equiv x_i(t-1)$, and	$F_{i}$ given by \eqref{SI:EQ:FKL}.
As a reminder, correlations remain between the observed states of node separated by more than 1 time step. Therefore, the likelihood is only approximately equal to the product of the transition probabilies [Eq.~\eqref{Eq:sis:likelihood}].

\section{Ecological networks}
The ecological networks are food webs downloaded from the Web of Life database (www.web-of-life.es). A summary of structural properties is presented in Table~\ref{SI:Table:weboflifes}. Nodes are species while the edges are observed interactions of predation.

\begin{table}
	\centering
	\caption[Summary properties of ecological networks]{Summary properties of ecological networks. Average, maximum, and minimum values of structural properties of the 25 food webs we have used. The average local clustering is the average of the fraction of existing over the possible triangles through each node \cite{newman2018networks}.}
	\label{SI:Table:weboflifes}
	\begin{tabular}{L{5cm}ccc}
		Network property & Mean & Min & Max\\
		\hline
		Nodes & 76.2 & 14 & 249\\
		Edges &  487.4 & 38& 3313\\
		Density & 0.097 &  0.0293 & 0.263\\
		Average local clustering & 0.10 & 0.016 & 0.358\\
		\hline
	\end{tabular}
\end{table}

\bibliographystyle{siam}
\bibliography{supp_refs}